\documentclass[12pt,preprint,natbib209]{aastex}

\usepackage{natbib}

\newcommand{\blob}{SST24 J1434110+331733}

\newcommand{\boo}{Bo\"otes}
\newcommand{\lya}{Ly$\alpha$}
\newcommand{\ltsim}{\raisebox{-0.6ex}{$\,\stackrel
        {\raisebox{-.2ex}{$\textstyle <$}}{\sim}\,$}}
\newcommand{\gtsim}{\raisebox{-0.6ex}{$\,\stackrel
        {\raisebox{-.2ex}{$\textstyle >$}}{\sim}\,$}}
\newcommand{\Msun}{M_\odot}
\newcommand{\Lsun}{L_\odot}
\newcommand{\Mdot}{\dot M}

\slugcomment{Accepted to ApJ 2005mar26}

\shorttitle{A Large Gaseous Nebula at z=2.7}
\shortauthors{Dey et al.}

\begin{document}

\title{Discovery of a Large $\sim$200~kpc Gaseous Nebula at $z\approx2.7$ with the {\it Spitzer Space Telescope}}

\author{Arjun Dey\altaffilmark{1}, 
Chao Bian\altaffilmark{2},
Baruch T. Soifer\altaffilmark{2,3}, 
Kate Brand\altaffilmark{1},
Michael J. I. Brown\altaffilmark{1,4},
Frederic H. Chaffee\altaffilmark{5},
Emeric LeFloc'h\altaffilmark{6},
Grant Hill\altaffilmark{5},
James R. Houck\altaffilmark{7},
Buell T.\ Jannuzi\altaffilmark{1}, 
Marcia Rieke\altaffilmark{6},
Daniel Weedman\altaffilmark{7},
Mark Brodwin\altaffilmark{8},
Peter Eisenhardt\altaffilmark{8}
}

\altaffiltext{1}{National Optical Astronomy Observatory, 950 N. Cherry Ave., Tucson, AZ 85719; dey@noao.edu}
\altaffiltext{2}{California Institute of Technology, Pasadena, CA 91125}
\altaffiltext{3}{Spitzer Science Center, California Institute of Technology, MS 220-6, Pasadena, CA 91125}
\altaffiltext{4}{Dept. of Astrophysical Sciences, Peyton Hall, Princeton University, Princeton, NJ 08544-1001}
\altaffiltext{5}{W. M. Keck Observatory, 65-1120 Mamalahoa Hwy., Kamuela, HI 96743}
\altaffiltext{6}{Steward Observatory, University of Arizona, Tucson, AZ 85721}
\altaffiltext{7}{Astronomy Department, Cornell University, Ithaca, NY 14853}
\altaffiltext{8}{Jet Propulsion Laboratory, California Institute of Technology, MC 169-327, 4800 Oak Grove Drive, Pasadena, CA 91109}

\begin{abstract}
We report the discovery of a very large, spatially extended \lya-emitting nebula at $z=2.656$ associated with a luminous mid-infrared source. The bright mid-infrared source ($F_{24\micron}=0.86$ mJy) was first detected in observations made using the Spitzer Space Telescope. Existing broad-band imaging data from the NOAO Deep Wide-Field Survey revealed the mid-infrared source to be associated with a diffuse, spatially extended, optical counterpart in the $B_W$ band.  Spectroscopy and further imaging of this target reveals that the optical source is an almost purely line-emitting nebula with little, if any, detectable diffuse continuum emission. The Ly$\alpha$ nebula has a luminosity of $L_{\rm Ly\alpha} \approx 1.7\times10^{44}{\rm \ erg\ s^{-1}}$ and an extent of at least 20\arcsec\ (160~kpc). Its central $\approx 8\arcsec$ shows an ordered, monotonic velocity profile; interpreted as rotation, this region encloses a mass $M\approx 6\times10^{12}\Msun$. Several sources lie within the nebula. The central region of the nebula shows narrow ($\approx 365\ {\rm km\ s^{-1}}$) emission lines of \ion{C}{4} and \ion{He}{2}. The mid-infrared source is a compact object lying within the nebula, but offset from the center by a projected distance of  $\approx 2\farcs5$ (20 kpc), and likely to be an  enshrouded AGN. A young star-forming galaxy lies near the northern end of the nebula. We suggest that the nebula is a site of recent multiple galaxy and AGN formation, with the spatial distribution of galaxies within the nebula perhaps tracking the formation history of the system.

\end{abstract}


\section{Introduction}

How do the most massive galaxies form? This is a question whose answer may provide one of the central tests of hierarchical galaxy formation theories. Our current theoretical paradigm suggests that the most massive galaxies form, like the rest of the galaxy population, by the hierarchical merging of smaller units and that this process occurs through the history of the universe. In seeming contrast, however, three lines of evidence suggest that at least some of the most massive galaxies may have had more dramatic origins. First, observations of the most massive galaxies at redshifts out to $z\ltsim 1.5$ suggest that the bulk of the stars in these objects formed at much higher redshifts in a brief time period characterized by very high star-formation rates ($\gtsim 300{\rm \Msun\,yr^{-1}}$) \citep[e.g.,][]{mcc2004}. Second, the host galaxies of powerful radio sources, which appear to be dynamically relaxed massive ellipticals at redshifts out to one \citep[e.g.,][]{zir2003}, at higher redshifts ($z\gtsim 3$) are surrounded by large (100$-$200~kpc), luminous ($>10^{43}{\rm erg\,s^{-1}}$) Ly$\alpha$ nebulae and small, embedded, star-forming objects \citep[e.g.,][]{reu2003}. This suggests that these host galaxies are formed in a rather spectacular gravitational collapse event rather than by the slow build-up of many merging events over time \citep[]{reu2003}. The discovery of large, diffuse \lya\ haloes in several surveys and their association with large galaxy overdensities adds further support to the idea that these haloes mark the formation sites of massive galaxies \citep[e.g.,][]{kee1999, ste2000, fra2001, pal2004, mat2004}. Third, the recent discoveries that dust-enshrouded galaxies selected in sub-mm surveys have very large star-formation rates, large correlation length and high median redshift suggest that this population may represent the formation epoch of massive galaxies \citep[e.g.,][]{bla2004,cha2004a}. Taken together, the empirical picture that emerges for massive galaxy formation is one of a violent, perhaps brief epoch of dust-enshrouded star-formation followed by a more gradual unveiling of the galaxy by the dissipation of its dusty, gaseous envelope. 

It is clearly of great interest, therefore, to find and characterize the formation sites of massive galaxies. Thus far we have been limited to studying the ($>$100~kpc) \lya\ halos and environments of luminous radio galaxies, which may be affected in peculiar ways by the presence of the active galactic nucleus and its ejecta, and to the similarly large \lya\ nebulae in two regions: four associated with a massive structure at $z=2.38$ \cite[]{fra2001,pal2004} and two others associated with a structure at $z=3.09$ \citep[]{ste2000,mat2004}. It is worth noting that the only narrow-band emission line surveys that have yielded discoveries of these spectacular \lya\ haloes have been targeted at large mass overdensities, so their mutual association may be misleading. Ground-based sub-mm surveys provide a less biased approach in principle, but so far have been limited by technology: the current suite of bolometer arrays provide poor positional accuracy and limited fields of view which are not suited to wide-field surveys. 

In this paper, we present observations that suggest that mid-IR surveys with {\it Spitzer Space Telescope} may provide an alternate approach: we report on observations of a luminous, very spatially extended \lya\ nebula at $z=2.6562$ discovered by virtue of its associated 24\micron\ emission. We present the description of our observations of this nebula in \S 2 and our results and measurements in \S 3. In \S 4, we discuss the implications of this discovery on future studies of massive galaxy formation.

Throughout this paper we use ${\rm H_0 = 70~km \  s^{-1} \  Mpc^{-1}}$,  ${\rm \Omega_m = 0.3}$, ${\rm \Omega_\lambda=0.7}$. At a redshift $z=2.6562$ the luminosity distance is $d_L = 21.946$~Gpc and 1~\arcsec\ corresponds to a physical size of 7.96~kpc.

\section{Observations}

The optical photometry of this source is derived primarily from the images obtained by the NOAO Deep Wide-Field Survey \citep[NDWFS\footnote{See also http://www.noao.edu/noaodeep/}; ][]{ndwfs, jan2005, dey2005}. The NDWFS is a deep, ground-based, optical and near-infrared imaging survey of two 9.3 square degree fields, one in \boo\ and one in Cetus, conducted using the 4m and 2.1m telescopes of the National Optical Astronomy Observatory. The survey is designed to study the evolution and clustering of galaxies over a wide range in redshift, and reaches median 5$\sigma$ point-source depths in the $B_W, R, I$ and $K$ bands of $\approx$27.1, 26.1, 25.4 and 19.0 mag (Vega) respectively. The data products for the NDWFS will be described elsewhere \citep[in preparation]{dey2005,jan2005}. The specific optical image stacks used in this paper have $B_W$, $R$, and $I$ band exposure times of 2.67, 2.67 and 3.33 hours and reach 5$\sigma$ point-source depths of 27.31, 26.23 and 25.47 mag respectively.  Stars in all three image stacks have FWHM$\approx$1\arcsec. The source is undetected in the near-infrared $K_S$ data of the NDWFS. The NDWFS astrometry is tied to the reference frame defined by stars from the USNO A-2 catalog \citep[]{usnoA2}. 

The NDWFS \boo\ field was mapped at 24, 70 and 160\micron\ using the Multiband Imaging Photometer for Spitzer  \citep[MIPS; ][]{rie2004} in late January 2004 to 1$\sigma$ rms depths of 51$\mu$Jy, 5mJy and 18mJy respectively. The data were reduced by the MIPS GTO team. Details of the survey, including the mapping strategy, data reduction and the resulting catalog, will be discussed elsewhere (LeFloc'h et al. 2005, in preparation). 
The entire \boo\ field was also mapped at 3.6, 4.5, 5.6 and 8.0\micron\ using the Infrared Array Camera \citep[IRAC; ][]{faz2004} on {\it Spitzer}; details of the IRAC observations may be found in \citet[]{eis2004}. Both the MIPS and IRAC data are astrometrically calibrated to the 2MASS astrometric frame. 

Initial optical identifications of the MIPS sources with 24$\mu$m flux densities brighter than 0.75mJy were done using the image catalogs from the NDWFS. The MIPS 24\micron\ sources are all unresolved at the 6\arcsec\ angular resolution of the {\it Spitzer Space Telescope}. The images of all MIPS sources with optical associations fainter than $I$=23 mag were visually inspected. A 24\micron\ source \blob\ with a flux density of 0.86mJy was found to lie within a faint, diffuse, spatially extended optical source with blue optical colors, which suggested that the colors may be the result of strong line emission in the $B_W$ band. 

\begin{figure*}
\begin{center}
\plottwo{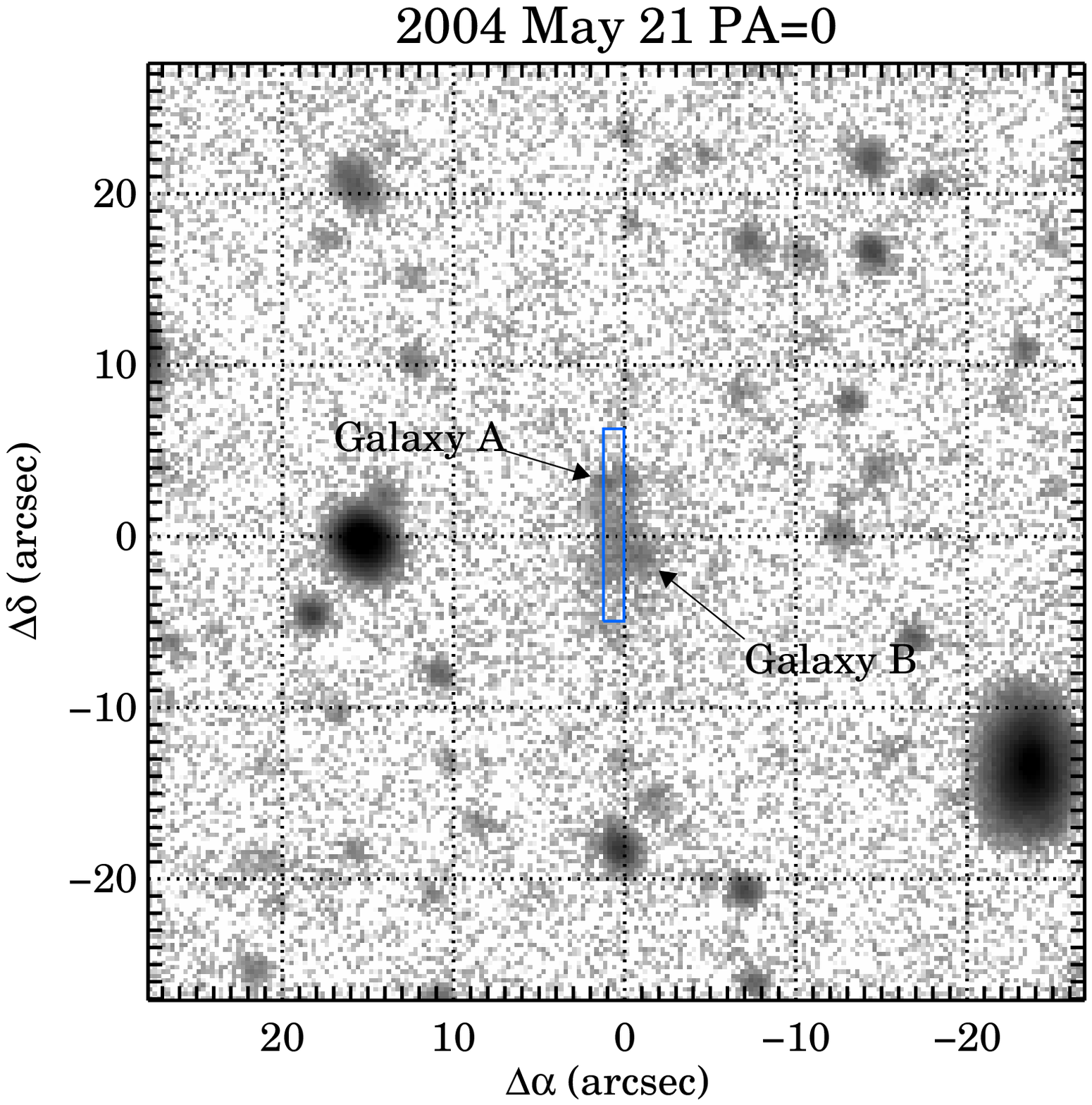}{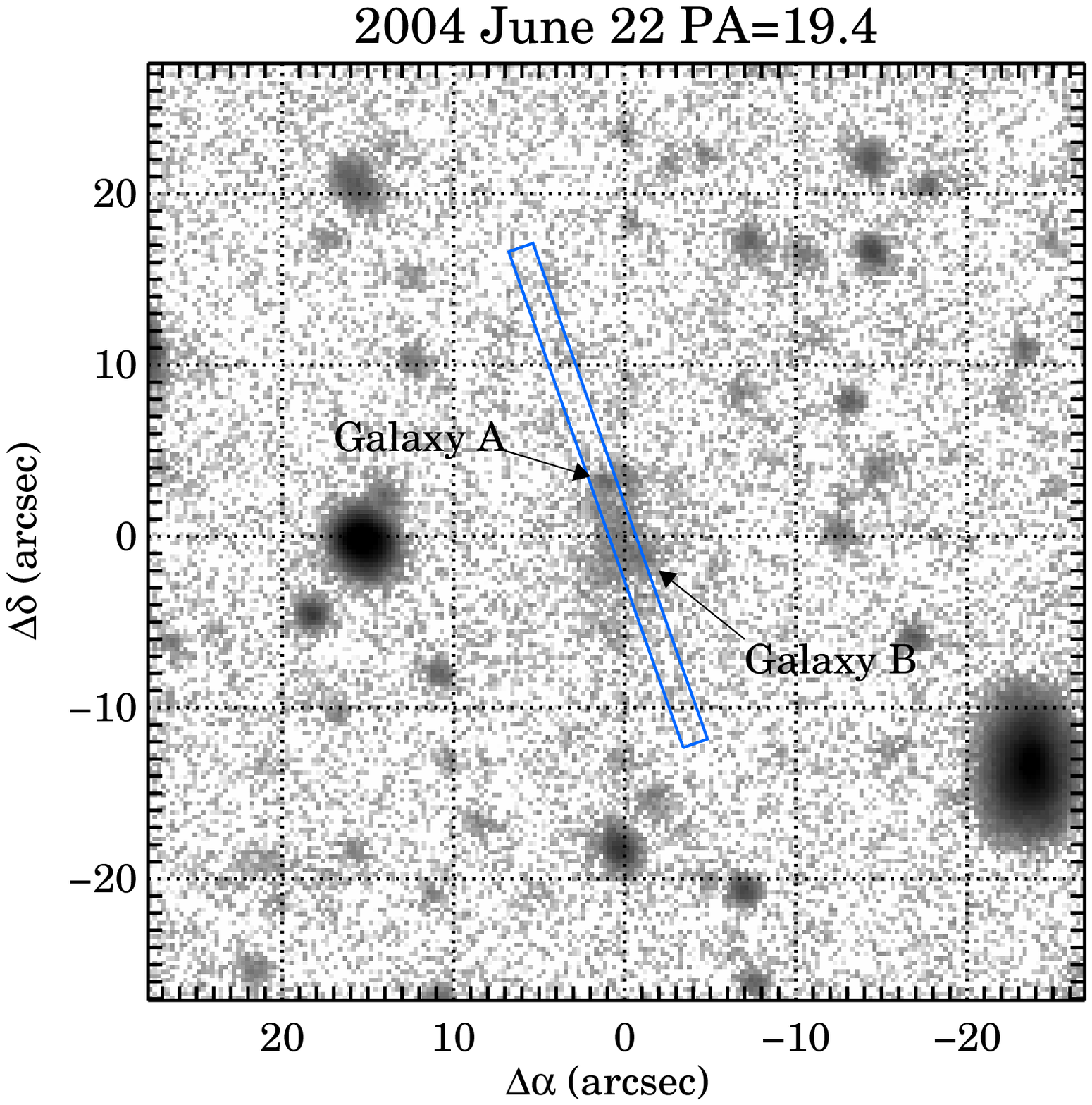}
\caption{The left and right panels show the location, orientation and size of the LRIS slitlets superimposed on the $B_W$ image of \blob. The origin corresponds to $14^h 34^m 10\fs977\ +33^\circ 17^\prime 30\farcs87$, the centroid of the \lya\ emission as determined from the narrow-band image (please see text). Note that only the slitlet at PA=19.4 (on the right) includes light from galaxy B.}
\label{specslits}
\end{center}
\end{figure*}

Spectroscopic observations of the MIPS 24\micron\ candidate \blob\  were obtained using the Low Resolution Imaging Spectrometer \citep[LRIS; ][]{oke1995} on the Keck I telescope on U.T. 2004 May 21 and 2004 June 22. The first set of observations, four 30~min exposures, were obtained in 0.8\arcsec\ seeing through a 1.0\arcsec\ wide and 11\arcsec\ long slitlet oriented at PA=$0^\circ$. LRIS was configured with the 5600\AA\ dichroic, the 400l/mm ($\lambda_c$=3400\AA, 1.09\AA\ pix$^{-1}$) grism on the blue side and the 400l/mm ($\lambda_b$=8500\AA, 1.86\AA\ pix$^{-1}$) grating on the red side. A second set of five 30~min observations were obtained in 1\arcsec\ seeing through a 1.5\arcsec\ wide and 30\arcsec\ long slitlet oriented at PA=$19.4^\circ$, with the 6800\AA\ dichroic, 300l/mm ($\lambda_c$=5000\AA, 1.43\AA\ pix$^{-1}$) grism, and the same red grating. In each observation set, the telescope was offset by a few arcsec between individual 30~min exposures. Relative spectrophotometric calibrations were performed using observations of Wolf~1346 \citep[]{oke1990,mas1988,mas1990}. Figure~\ref{specslits} shows the location and orientation of the LRIS slitlets on the $B_W$ image of the source. 

Additional optical observations of the field were obtained using the MiniMOSAIC camera (``MiniMo'') on the WIYN 3.5-m telescope of the Kitt Peak National Observatory on the nights of U.T. 2004 June 14 -- 17. A total of six 1200~s images were obtained through the 62\AA-wide narrow-band filter centered at 4448\AA\ (hereinafter NB4448, also called ``W20'' in the WIYN filter set; centered on the redshifted \lya\ line) and three 900~s images were obtained through a Gunn $G$ filter. The seeing varied from night to night, but point sources in the final NB4448 and G-band  image stacks have an effective FWHM of 0\farcs7 and 0\farcs8 respectively.  The data were calibrated using observations of the spectrophotometric standard Feige 110 \citep[]{oke1990,sto1977} through the NB4448 filter and Landolt field SA112 \citep[]{lan1992} through the Gunn $G$ filter on photometric nights. 

All the WIYN optical imaging data were astrometrically calibrated using the USNO A-2 catalog and tied to the same astrometric frame as the NDWFS imaging data. There is a small astrometric offset between the IRAC and MIPS 2MASS frame and the NDWFS/USNO A-2 frame of roughly 0\farcs5 which we corrected before comparing the Spitzer and ground-based data. All positions quoted here are on the NDWFS/USNO A-2 reference frame.

\section{Results}

\subsection{Morphology and Astrometry}

\begin{figure*}
\plotone{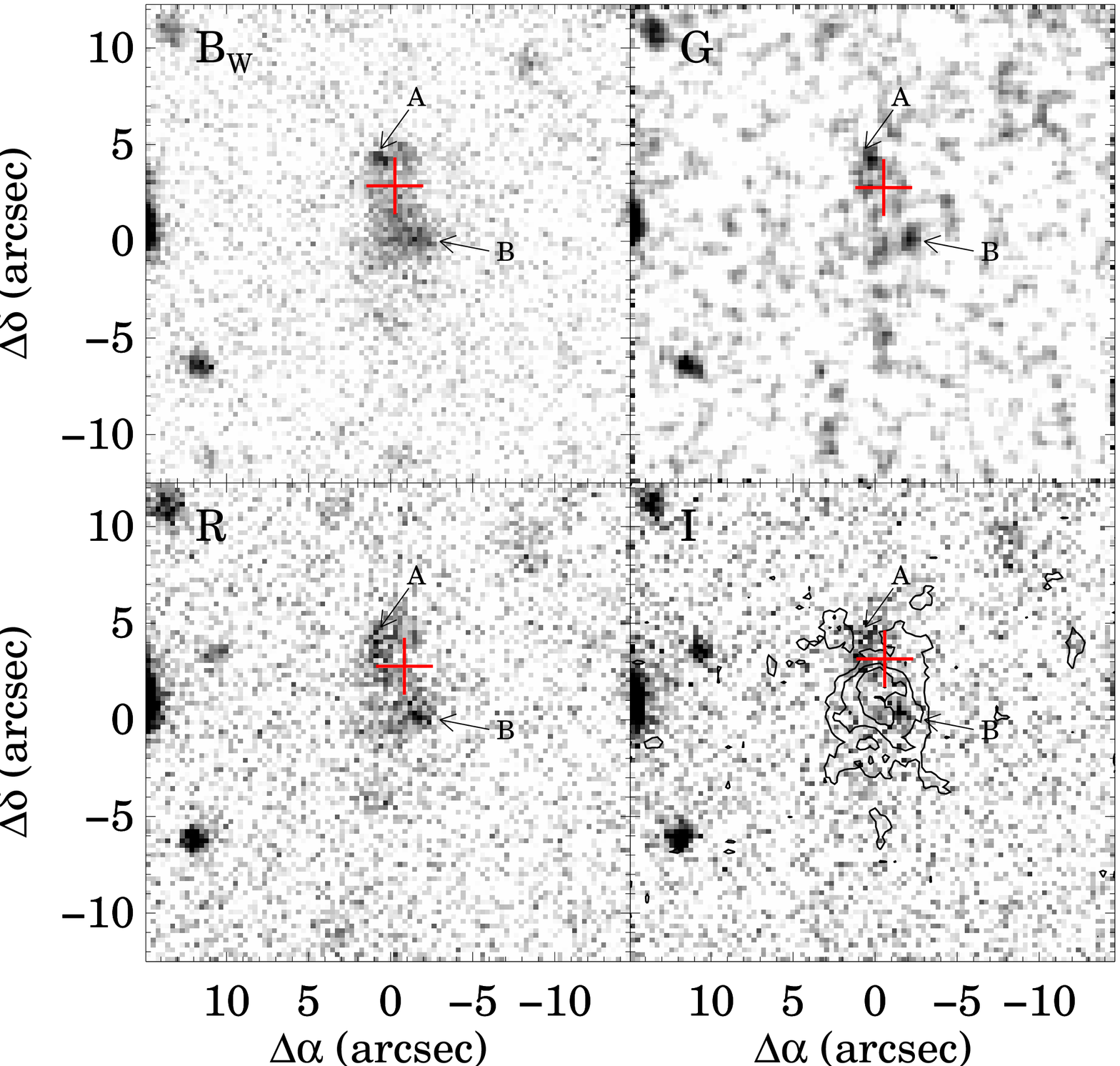}
\caption{Image cutouts of the \blob\ in $B_W$-, $G$-, $R$- and $I$-bands. The origin (0,0) in all panels is at $14^h34^m11\fs005 \ +33^\circ17^\prime29\farcs84$ (J2000). The red crosses in the four panels mark the location of the 24\micron\ source ($B_W$ panel), the 3.6\micron\ source ($G$ panel), the IRAC 4.5\micron\ source ($R$ panel) and the 8.0\micron\ source ($I$ panel). The position of the source in the MIPS 24\micron\ and IRAC bands agrees to $<0\farcs5$. The contours in the $I$ panel denote the emission detected in the NB4448 filter centered on the Ly$\alpha$ emission line. The continuum emission in this filter has not been subtracted, but contributes less than 10\% of the total flux at all radii. The contours drawn correspond to line flux surface brightnesses of approximately [0.2,0.4,0.6,0.8]$\times 10^{-16}{\rm erg\ s^{-1}\ cm^{-2}\ arcsec^{-2}}$.  \label{images}  }
\end{figure*}

Image cutouts of the $B_W$, $G$, $R$ and $I$ images are shown in figure~\ref{images}, with the contour map from the NB4448 image overlaid on the $I$-band image\footnote{Larger finding charts of the field and astrometry of offset stars may be determined from image cutouts of the NDWFS available from the NOAO Science Archive (http://www.archive.noao.edu/ndwfs/).}. The system is diffuse and relatively bright in the $B_W$ band, which is largely dominated by the \lya\ line emission; there is a fairly compact galaxy at its northern extent which is labelled `A' in figure~\ref{images}.  In contrast to the $B_W$ image, the $G$, $R$ and $I$ band cutouts show only little (if any) diffuse emission, but do show some light associated with the regions `A' and `B'. The \lya\ emission appears to avoid `A'. The center of the \lya\ emission lies near `B', but slightly to its east; the spectrum of this central region is therefore contaminated by continuum emission from `B'.

The NDWFS astrometry is tied to a frame defined by moderately bright (12$-$17 mag) stars in the USNO A-2 catalog, whereas the MIPS astrometry is calibrated using 2MASS. Roughly 98\% of the bright ($F_{24\micron}>0.75$ mJy) MIPS sources have well determined optical counterparts in the NDWFS data, and comparisons of the relative astrometry between these MIPS and NDWFS matches show a small, but significant relative offset of $\Delta\alpha=0{\farcs}38$ and $\Delta\delta=0\farcs14$ between the two frames. The offset arises because of a small magnitude-dependent astrometric error internal to the USNO A-2 catalog \citep[]{mon2003}. The MIPS source is located at $14^h34^m10\fs958\ +33^\circ17^\prime32\farcs57$ in the J2000 2MASS reference frame. In the optical reference frame of the NDWFS, the source position is $14^h34^m10\fs988\ +33^\circ17^\prime32\farcs71$  (J2000). Taking this offset into account, the MIPS source is found to lie on the northern end of the diffuse emission observed in the NDWFS $B_W$ and NB4448 images, just to the southwest of galaxy A (fig~\ref{images}). The location of the MIPS source is marked by a cross, and is believed to be accurate to $\approx 0\farcs5$. It lies $\approx 0\farcs8$ west and $1\farcs6$ south of galaxy A and $\approx 1\farcs1$ east and 2\farcs6 north of galaxy B. The \lya\ emission is centered roughly 2\farcs5, or 20~kpc (projected), away from the MIPS source. 

The position of the IRAC source detected in the 3.6\micron, 4.5\micron, and 8\micron\ bands agree with the MIPS 24\micron\ source position to $<$0\farcs5, and confirm the relative (non-central) location of the mid-infrared source within the \lya\ nebula. 

\subsection{Photometry}

Table~\ref{photom} presents our photometric data on \blob. The optical photometry was measured in a large aperture of radius 10\arcsec\ centered on the centroid of the \lya\ halo. The flux in the $B_W$ band is dominated by the \lya\ emission, and the optical measurement aperture for all bands includes galaxies A and B. These optical measurements should therefore be strictly treated as upper limits to the continuum emission from the region. Higher spatial resolution imaging and spectroscopy along PA$\approx$90$^\circ$ will be necessary to disentangle the contribution of galaxy B and determine the true continuum contribution from sources within the nebula.

\begin{figure*}
\begin{center}
\plotone{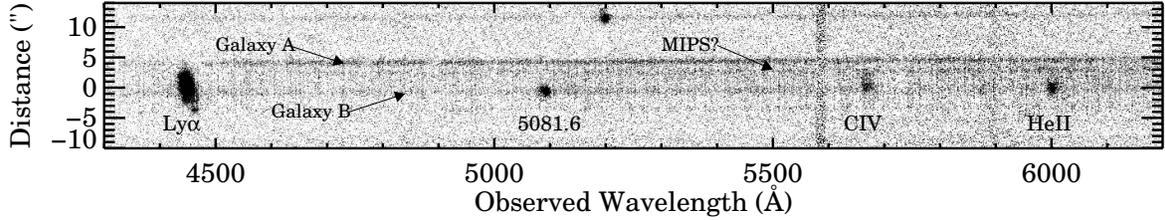}
\caption{Two-dimensional spectrum of the \lya\ nebula associated with \blob\ obtained through a 1\farcs5 wide slitlet oriented in PA=19.4$^\circ$ and using LRIS-B on the Keck I telescope. The three continuum streaks and main emission lines are labelled. Please see text for details.}
\label{spec2d}
\end{center}
\end{figure*}

\begin{deluxetable}{lcccccc}
\tabletypesize{\scriptsize}
\tablecaption{Photometry of \blob\ \label{photom}}
\tablewidth{0pt}
\tablehead{
\colhead{Band} & \colhead{Wavelength} & \colhead{Width} & \colhead{Aperture Center} & \colhead{Aperture} & \colhead{Magnitude} & \colhead{$F_\nu$} \\
      &  \colhead{(\micron)} & \colhead{(\micron)} &   \colhead{(J2000)} & \colhead{Radius} & \colhead{(Vega)} & \colhead{(${\rm erg~s^{-1}~cm^{-2}~Hz^{-1}}$)} 
}
\startdata
NDWFS $B_W$ & 0.422 & 0.112 & $14^h 34^m 10\fs977\  +33^\circ 17^\prime 30\farcs87$ & 10\arcsec & $22.64\pm0.05$& $3.18\pm0.15 \times 10^{-29}$ \\
Gunn $G$           & 0.492 & 0.067 &  " & 10\arcsec & $23.02\pm0.20$ & $2.39\pm0.45 \times 10^{-29}$\\
NDWFS $R$      & 0.659 & 0.183 &  " & 10\arcsec & $22.24\pm0.08$ & $3.92\pm0.30\times 10^{-29}$ \\
NDWFS $I$        & 0.808 & 0.172 &  " & 10\arcsec & $22.02\pm0.15$ & $3.80\pm0.53\times 10^{-29}$ \\
NDWFS $K_S$  & 2.167 & 0.295 & " & 6\arcsec & $>18.12$\tablenotemark{a}  & $<3.67\times 10^{-28}$\tablenotemark{a}  \\
IRAC 3.6\micron & 3.56   & 0.75  & $14^h 34^m 10\fs979\ +33^\circ 17^\prime 32\farcs48$ & 6\arcsec & $17.83\pm0.13$  & $2.04\pm0.24\times 10^{-28}$\\
IRAC 4.5\micron & 4.52   & 1.01  & " & 6\arcsec & $17.09\pm0.13$  & $2.63\pm0.31\times 10^{-28}$\\
IRAC 5.8\micron & 5.73   & 1.42  & " & 6\arcsec  & $15.93\pm0.36$  & $4.97\pm1.66\times10^{-28}$ \\
IRAC 8\micron    & 7.91    & 2.93  & " &  6\arcsec &$14.79\pm0.21$ & $7.69\pm1.49\times 10^{-28}$\\
MIPS 24\micron  & 24.0   & 5.3  & $14^h 34^m 10\fs988\ +33^\circ 17^\prime 32\farcs71$ & \tablenotemark{b} & $9.83\pm0.06$ & $8.56\pm0.51\times 10^{-27}$ \\
MIPS 70\micron  & 70.0   & 19.0 & & & & $<2.5\times10^{-25}$ \\
MIPS 160\micron&160.0 & 34.5 & &  & & $<9.0\times 10^{-25}$ \\
WSRT 1400MHz\tablenotemark{c}  & 20cm &        & $14^h 34^m 10\fs78\ +33^\circ 17^\prime 27\farcs4$ &   &   & $1.4\pm0.4\times10^{-27}$ \\
\enddata
\tablenotetext{a}{The $K_S$ upper limit is 2$\sigma$ in a 6\arcsec\ radius aperture.}
\tablenotetext{b}{The MIPS photometry was determined using PSF-fitting techniques.}
\tablenotetext{c}{WSRT measurement is from the deep 1400MHz survey of \citet[]{dev2002}. The source is unresolved in the 13\arcsec$\times$27\arcsec\ beam. }
\end{deluxetable}

\subsection{Redshifts of the Nebula and Coincident Sources}

\begin{figure*}
\begin{center}
\plotone{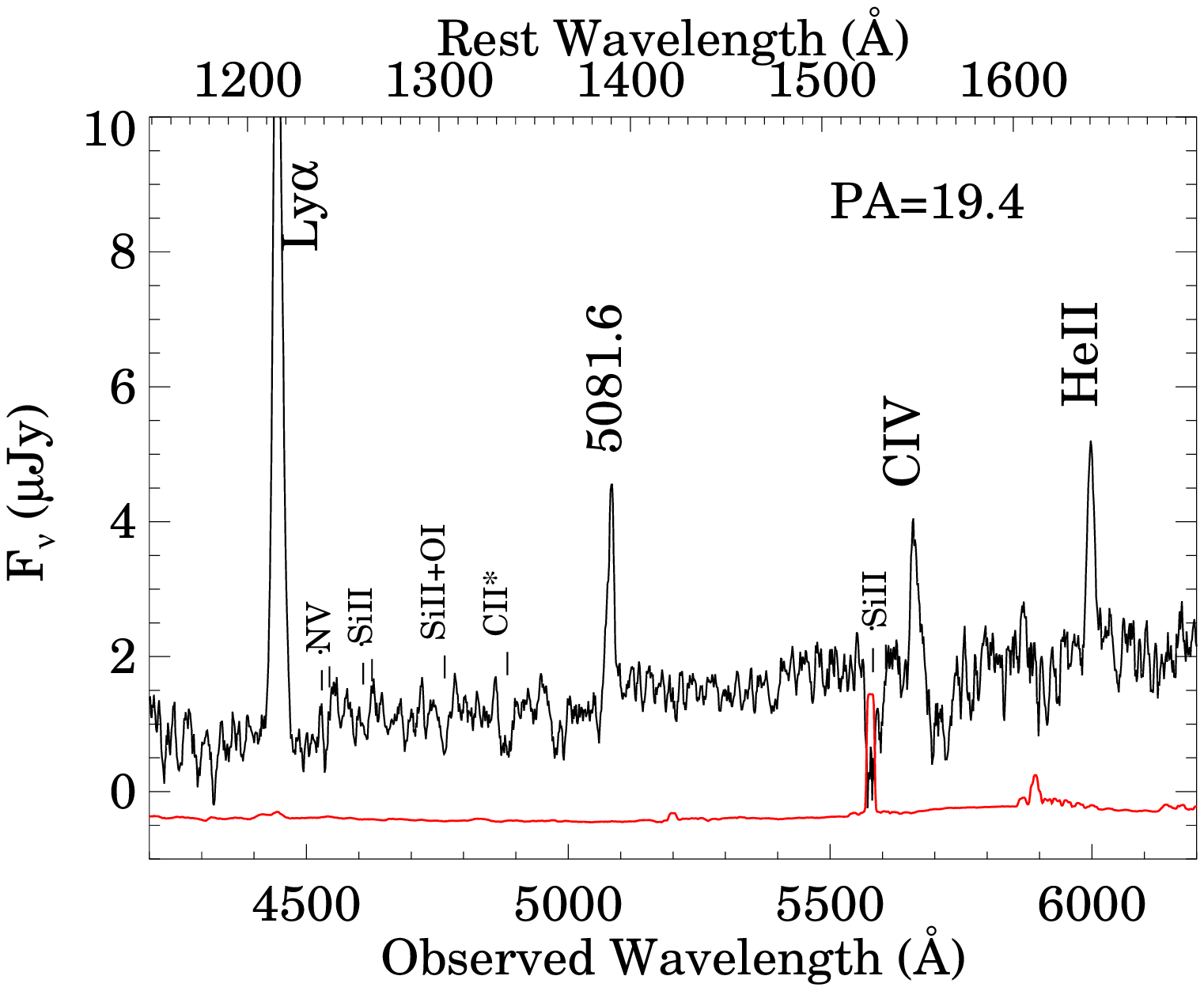}
\caption{One-dimensional spectrum of the \lya\ nebula associated with \blob\ smoothed by 10\AA. The aperture is 1\farcs5 wide and 10\farcs9 long and oriented at PA=19.4$^\circ$. The red line shows the noise spectrum offset by $-1\mu$Jy. The absorption lines arise predominantly from galaxy A, with some contribution from galaxy B noticeable at $\lambda>5100$\AA. The interloping emission line at $\lambda$=5081.6\AA\ is only observed in the PA=19.4$^\circ$ observation, and is therefore thought to arise from galaxy B, which lies just to the west of the emission center of the \ion{C}{4} and \ion{He}{2} line emission.}
\label{spec1d}
\end{center}
\end{figure*}

The spectroscopic observations reveal a very complex system within the extent of the nebula, comprised of several continuum sources. Figure~\ref{spec2d} shows the two-dimensional spectrum recorded on the blue side of LRIS in PA=19.4$^\circ$, and figure~\ref{spec1d} shows the extracted one-dimensional spectrum of the system obtained through a (wide) 10\farcs9$\times$1\farcs5 aperture from these data. Emission line measurements are presented in table~\ref{specdata}. The redshift of the system is well established at $z=2.6562$ through the detection of \lya, \ion{C}{4}, \ion{He}{2} and weak \ion{C}{3}]. 

\begin{figure*}
\begin{center}
\plotone{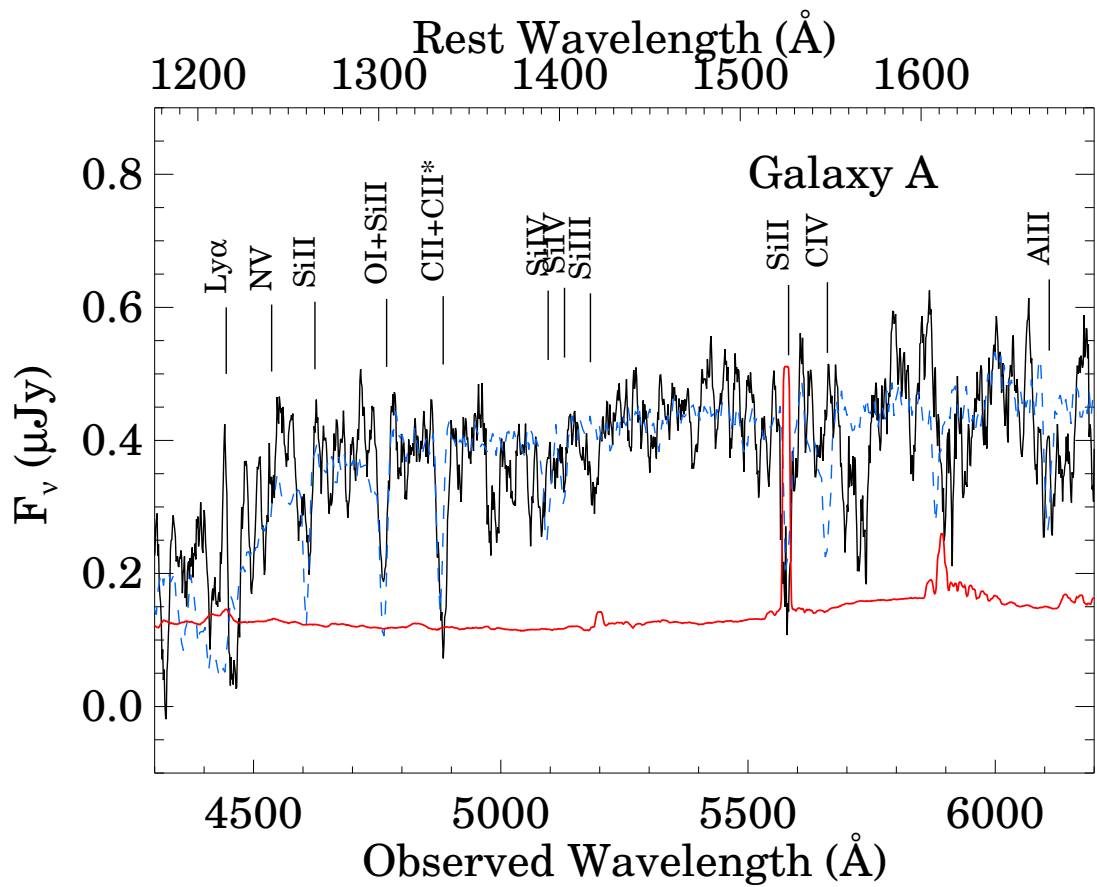}
\caption{One-dimensional spectrum of Galaxy A extracted in an aperture 1\farcs5 $\times$ 0\farcs95 from the PA=19.4$^\circ$ observation. The dashed line overlaid on the spectrum is the composite spectrum of a $z\sim 3$ star-forming galaxy constructed from a sample showing strong \lya\ absorption \cite[]{sha2003}.}
\label{spec1d_galA}
\end{center}
\end{figure*}

\begin{figure*}
\begin{center}
\plotone{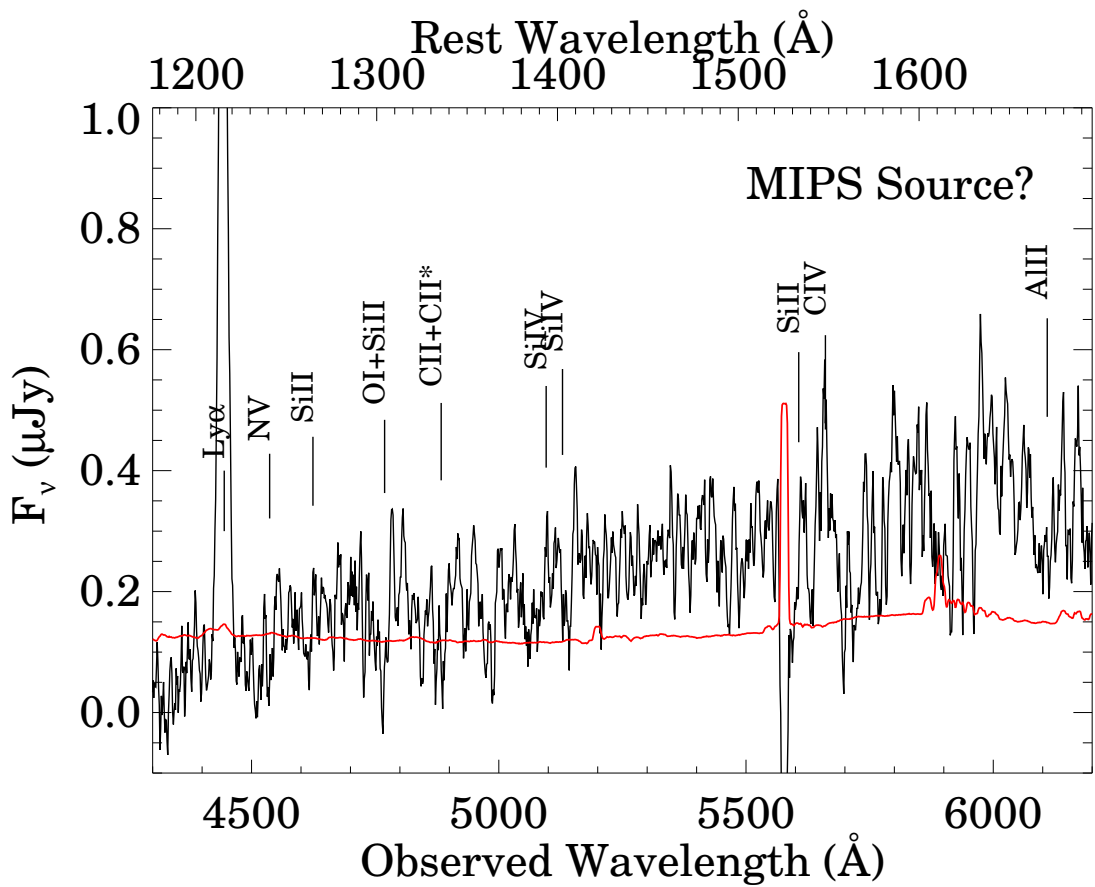}
\caption{One-dimensional spectrum of the continuum object closest to the location of the mid-infrared source. The extraction aperture is 1\farcs5 $\times$ 0\farcs95 from PA=19.4$^\circ$ observation. The spectrum is much redder than that of galaxy A (figure~\ref{spec1d_galA}) but shows weak \ion{C}{4}.}
\label{spec1d_mips}
\end{center}
\end{figure*}

\begin{figure*}
\begin{center}
\plotone{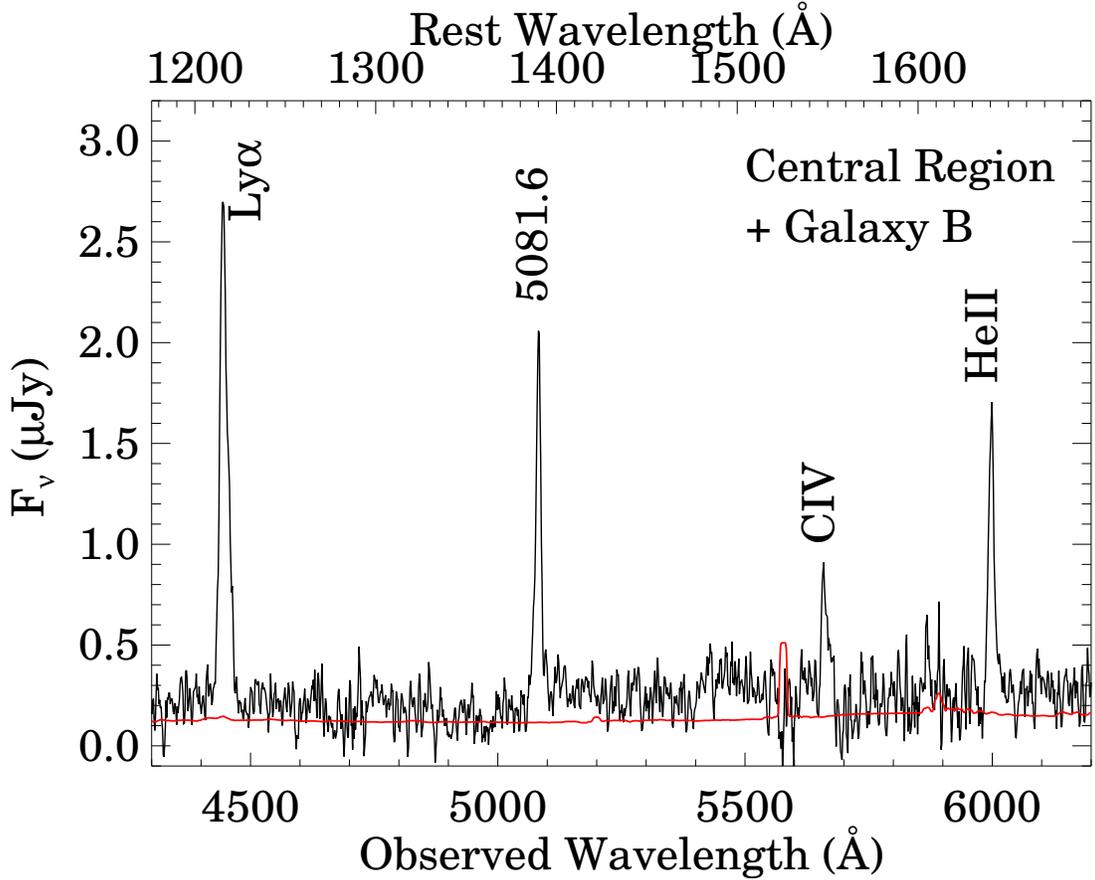}
\caption{One-dimensional spectrum of central region and galaxy B from the PA=19.4$^\circ$ observation. The extraction aperture is 1\farcs5 $\times$ 0\farcs95. The spectrum cannot be disentangled cleanly since the two contributing components lie parallel to the dispersion direction, but we suggest that the 5081.6\AA\ line and the continuum emission are dominated by galaxy B, whereas the \lya, \ion{C}{4} and \ion{He}{2} emission arise from the central region of the nebula.}
\label{spec1d_cen}
\end{center}
\end{figure*}

There are three main continuum contributors to this large-aperture spectrum: (i) galaxy A (figure~\ref{spec1d_galA}), (ii) a source 1\farcs6 southwest of galaxy A along the PA=19.4$^\circ$ slitlet which may be associated with the MIPS source (figure~\ref{spec1d_mips}), and (iii) a source 4\farcs7 southwest of A, and close to the location of galaxy B (figure~\ref{spec1d_cen}). We discuss each of these in turn. 

The spectrum of galaxy A (figure~\ref{spec1d_galA}) reveals a ``Lyman-break galaxy'' at a redshift of $z\approx 2.656$, characterized by a flat $F_\nu$ continuum, interstellar absorption lines and no detectable \lya\ emission. The interstellar absorption lines observed in typical Lyman-break galaxies can be blue shifted due to winds, and the redshift estimate based on these is therefore only good to $\sim1000{\rm \ km\ s^{-1}}$. The absorption line spectrum seen in figure~\ref{spec1d} is almost entirely due to the continuum contribution from galaxy A to the large aperture extraction. 

The spectrum of the second continuum source (figure~\ref{spec1d_mips}) is quite red and shows weak \ion{C}{4} and \ion{C}{3}] emission at $z=2.656$, i.e., the redshift of the nebula. Its proximity to the MIPS 24\micron\ location suggests that it may be a star-forming galaxy reddened by dust. There are some possible continuum features visible in the spectrum of this source, that are consistent with a young, reddened star-forming object. No obvious emission lines uniquely due to an AGN are detected in this source. 

The spectra of galaxy A and the MIPS source both show a fairly broad absorption feature at $\lambda$5720\AA. The signal-to-noise ratio in this region is low, and sky subtraction using traditional fitting techniques on very extended faint objects can sometimes result in such features. If this feature is real, it is most likely to be due to a foreground absorption system, perhaps arising from either MgII$\lambda\lambda$2800 at $z\approx1.04$ or an FeII 2600\AA\ absorption complex at $z\sim1.20$."

The third continuum source (figure~\ref{spec1d_cen}) is the most intriguing: it lies near the center of the \lya\ nebula and apparently shows strong associated \ion{C}{4} and \ion{He}{2} emission at the redshift of the nebula. This continuum source also shows an emission feature at 5081.6\AA, the identification of which is uncertain. This feature is unlikely to be an emission line at the redshift of the \lya\ nebula, since it does not match any strong features normally observed in high-redshift galaxies or AGN: at a rest wavelength of 1389.9\AA, it could only be associated with blue-shifted SiIV$\lambda\lambda$1393.8,1402.8 emission, and would be the only feature with such a blue shift relative to the \ion{He}{2} redshift, which seems very improbable. This spectral line also does not appear in the PA=0$^\circ$ spectrum, strongly suggesting that it is associated with galaxy B. Since the 5081.6\AA\ line does not correspond to any normal species at the redshift of the \lya\ nebula, we therefore suspect that galaxy B is an interloping system that does not lie at the redshift of the nebula. Given the asymmetry of the 5081.6\AA\ emission line and the break in the continuum across the line, we suggest that galaxy B is a background source at a redshift of 3.180. Galaxy B is located too far away from the centroid of the MIPS 24\micron\ emission and is therefore also not the source of the infrared emission.  The spatial coincidence of the 5081.6\AA\ feature and the \ion{He}{2} and \ion{C}{4} emisson lines in our spectrum must therefore be due to the unfortunate circumstance of the two contributing sources, the emission line source at $z=2.656$ and the background galaxy at $z=3.18$, being located approximately along the short-axis of our slit, parallel to the dispersion direction. It is also likely that the continuum emission that we observe at this location is primarily due to galaxy B, and unrelated to the nebula. 

The source near the top of the slit in figure~\ref{spec2d} shows an asymmetric emission line at 5188.5\AA\ and a break across the line; this emission line is consistent with \lya\  at a redshift of 3.268, and this galaxy is also a background source. 

We are left, therefore, with the interpretation that two unrelated background sources are present in this 1\farcs5$\times$30\arcsec\ slitlet, both at redshifts above three.  We are confident the emission line detections associated with these objects are not instrumental effects, since they repeat on all the individual exposures and move in concert with their associated continuum when the targets were offset along the slit.  It is not exceptional to find a serendipitous source with $z>3$ during LRIS spectroscopy; our experience is that one such source appears in about 15\% of the masks used. Given the surface density of $z\sim 3$ Lyman break galaxies ($\approx$1.2 per sq. arcmin for $R\le 25.5$ mag; \cite[]{stei2003}), the Poisson probability of finding 2 in a single mask is 1.8\%, and in a single slitlet is 0.01\%. Note that these are lower limits to the probabilities since $z\sim3$ Lyman break galaxies are clustered. The redshift separation between the two background galaxies suggests a minimum comoving relative distance of $\approx$80~Mpc. Further spectroscopy of this region is clearly of interest in order to understand the field galaxy redshift distribution.

\begin{deluxetable}{lcccccc}
\tabletypesize{\scriptsize}
\tablecaption{Emission-Line Strengths\tablenotemark{a}
\label{specdata}}
\tablewidth{0pt}
\tablehead{
\colhead{Line} & \colhead{Obs. Wavelength} & \colhead{Aperture} & \colhead{Flux} & \colhead{FWHM} & \colhead{Obs. EW} & \colhead{Luminosity\tablenotemark{b}} \\
         & \colhead{(\AA)}  & \colhead{(\arcsec)} & \colhead{(${\rm erg\ s^{-1}\ cm^{-2}}$)} & \colhead{(\AA)} & \colhead{(\AA)} & \colhead{(${\rm erg\ s^{-1}}$)}
         }
\startdata
\lya  & 4448    & 10.0 radius\tablenotemark{c} & $2.89\pm0.15\times 10^{-15}$ & - & - & $1.67\times10^{44}$ \\
\\
\lya  & 4444.7 & 10.9$\times$1.5 & $4.05\pm0.01\times10^{-16}$ & 13.0   & 316 & $2.33\times10^{43}$ \\
\\
\lya  & 4445.0 & 4.5$\times$1.5 & $3.10\pm0.01\times10^{-16}$ &  12.8    & 396 & $1.79\times10^{43}$ \\
NV   &   -           & 4.5$\times$1.5 & $<5.1\times10^{-18}$ \tablenotemark{d} & - &  $<$7.3  & \\
\ion{C}{4}  & 5662.0 & 4.5$\times$1.5 & $4.17\pm0.04\times10^{-17}$ & 15.2\tablenotemark{e} &  85 & $2.40\times10^{42}$ \\
\ion{He}{2} & 5997.9 & 4.5$\times$1.5 & $4.07\pm0.04\times10^{-17}$ &  7.3      & 60 & $2.35\times10^{42}$ \\
\ion{C}{3}] & 6974.6 & 4.5$\times$1.5 & $5\pm1\times10^{-18}$             & $<$13 &    & $4.6\times10^{41}$ \\
    -   & 5081.6 & 4.5$\times$1.5 & $5.15\pm0.04\times10^{-17}$ & $<$4    & 87 & \\ 
\enddata
\tablenotetext{a}{All quantities are measured from the PA=19.4$^\circ$ spectrum with the exception of the first entry in the table, which is measured from the narrow-band image.}
\tablenotetext{b}{\ \ Luminosities are calculated for a cosmology with ${\rm H_0 = 70~km\ s^{-1}\ Mpc^{-1}, \ \Omega_m = 0.3, \ \Omega_\lambda = 0.7}$.}
\tablenotetext{c}{Measured from the narrow-band image after subtracting the continuum flux estimated from the G-band image}
\tablenotetext{d}{Upper limits quoted are 3$\sigma$ and assume lines of the same width as the \ion{He}{2} feature.}
\tablenotetext{e}{\ion{C}{4} width is measured by modelling the doublet as a single feature.}
\end{deluxetable}

\subsection{Spatial Extent of the \lya\ Nebula}

\begin{figure*}
\plotone{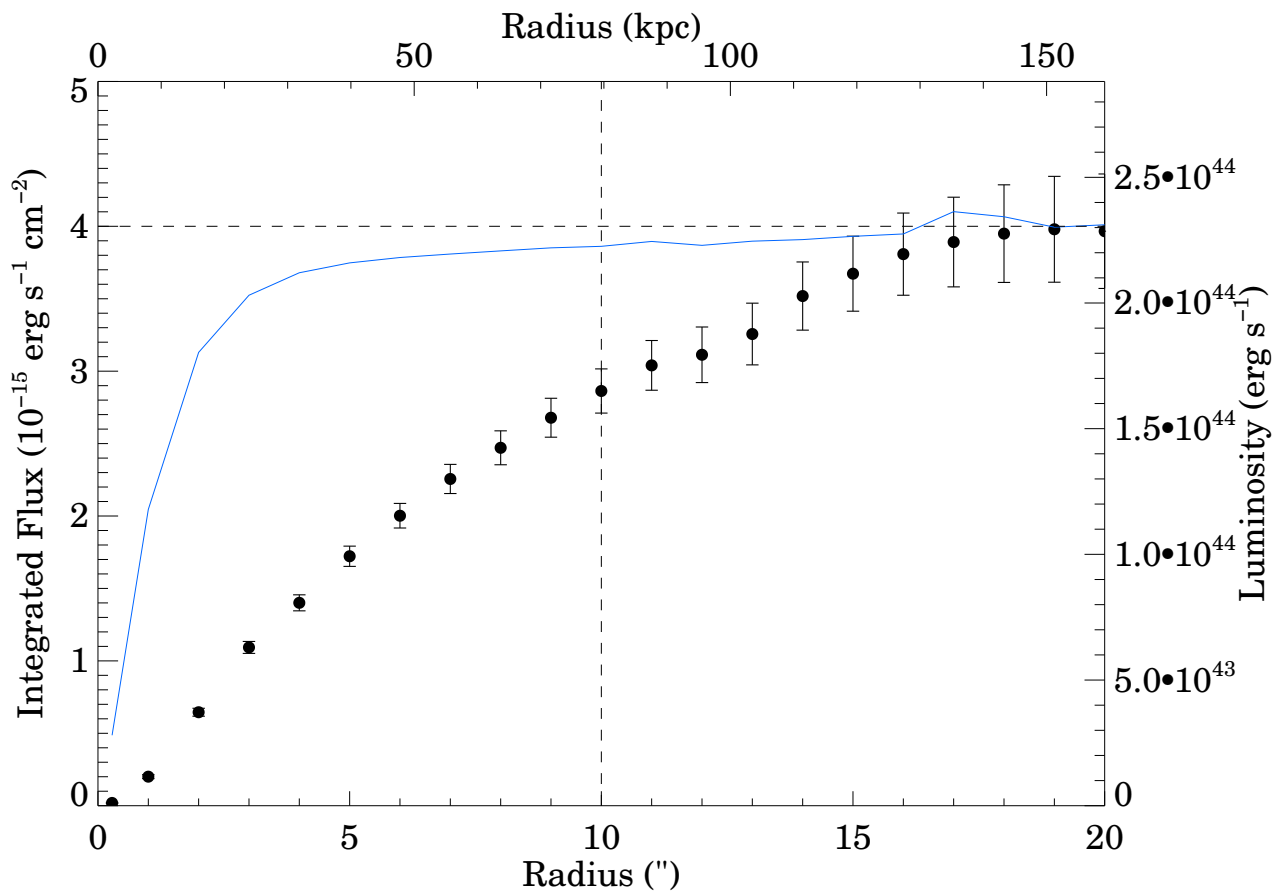}
\caption{The integrated flux of \lya\ as a function of radius as measured in the NB4448 image (filled circles). The solid line shows the mean stellar profile in the image, determined using 10 stars. The filled circles represent the integrated \lya\ profile after subtraction of the continuum estimated using the Gunn $G$-band observations. The dashed vertical line at a radius of 10\arcsec\ denotes our chosen (conservative) aperture size for most of the measurements presented in this paper. The horizontal dashed line is for reference. \label{profile}}
\end{figure*}

A visual examination of the NB4448 image at a hard stretch shows that the high surface brightness portion of the \lya\ nebula is at least 10\arcsec\ in size, but that there is faint diffuse emission extending over larger scales. In order to explore the full spatial extent of the nebula, we compared its radial profile with that of stars measured in the same image. Figure~\ref{profile} shows the integrated radial profile of the nebula compared with the median profile determined from 10 stars in the NB4448 image, normalized to the same total flux. It is clear that the nebula is spatially extended to a {\it radius} of at least 15\arcsec\ (nearly 120~kpc)! There are hints that the nebula may be even more extended, but deeper narrow-band imaging observations are necessary to confirm this.

The emission seen in the narrow-band filter is not significantly affected by continuum emission.
The Gunn $G$ filter only spans the wavelength range 1240$-$1460\AA\ in the rest frame of the system, and therefore excludes the redshifted \lya, \ion{C}{4} and \ion{He}{2} line emission from the nebula. This band provides a good measurement of the line-free flux from the region which we can use to estimate (and subtract) the continuum contribution in the NB4448 filter. In an aperture of 10\arcsec\ radius, the continuum contribution is only $0.225\times 10^{-15}\ {\rm erg\ s^{-1}\ cm^{-2}}$ in the NB4448 filter, i.e., only 7\% of the total flux. This estimate assumes that the continuum components are due to interlopers and contribute across the entire wavelength region of the narrow-band filter. If we instead assume that the continuum arises from sources at the redshift of the nebula, then the continuum contribution within the narrow-band filter would be reduced due to \lya\ forest absorption in the intergalactic medium. Using the Madau (1995) approximation for the total continuum absorption, we estimate that the continuum contribution would be about 13\% smaller under this assumption, and the total derived \lya\ flux imperceptibly larger.

\subsection{Velocity Structure in the Nebula}

Our spectroscopic observations are not sufficiently deep to detect the nebula across its full extent, and we restrict our discussion here to the central 8\arcsec\ region. 

\lya\ is a resonance line, and its profile is generally dominated by radiative transfer effects, being affected by resonant scattering and absorption by gas and dust. It is therefore generally a poor tracer of the gravitational potential or the physical properties of the gas. However, at present, this is the only real tracer we have of this system; we therefore use it to provide some indication of the kinematic properties of the system with these caveats in mind. 

We extracted spectra in 0\farcs4 (3-pixel) apertures from each of our spectroscopic observations and fit the \lya\ profile with a Gaussian model. The fits are fairly good approximations to the emission line profile; the profiles are only very slightly asymmetric across most of the nebula. We also performed similar fits on the [OI]$\lambda$5577 night sky emission line, and found that the tilt this line varied by $\le \pm5\ {\rm km\ s^{-1}}$ across the portion of the slitlets covering the \lya\ nebula. 

\begin{figure*}
\begin{center}
\plotone{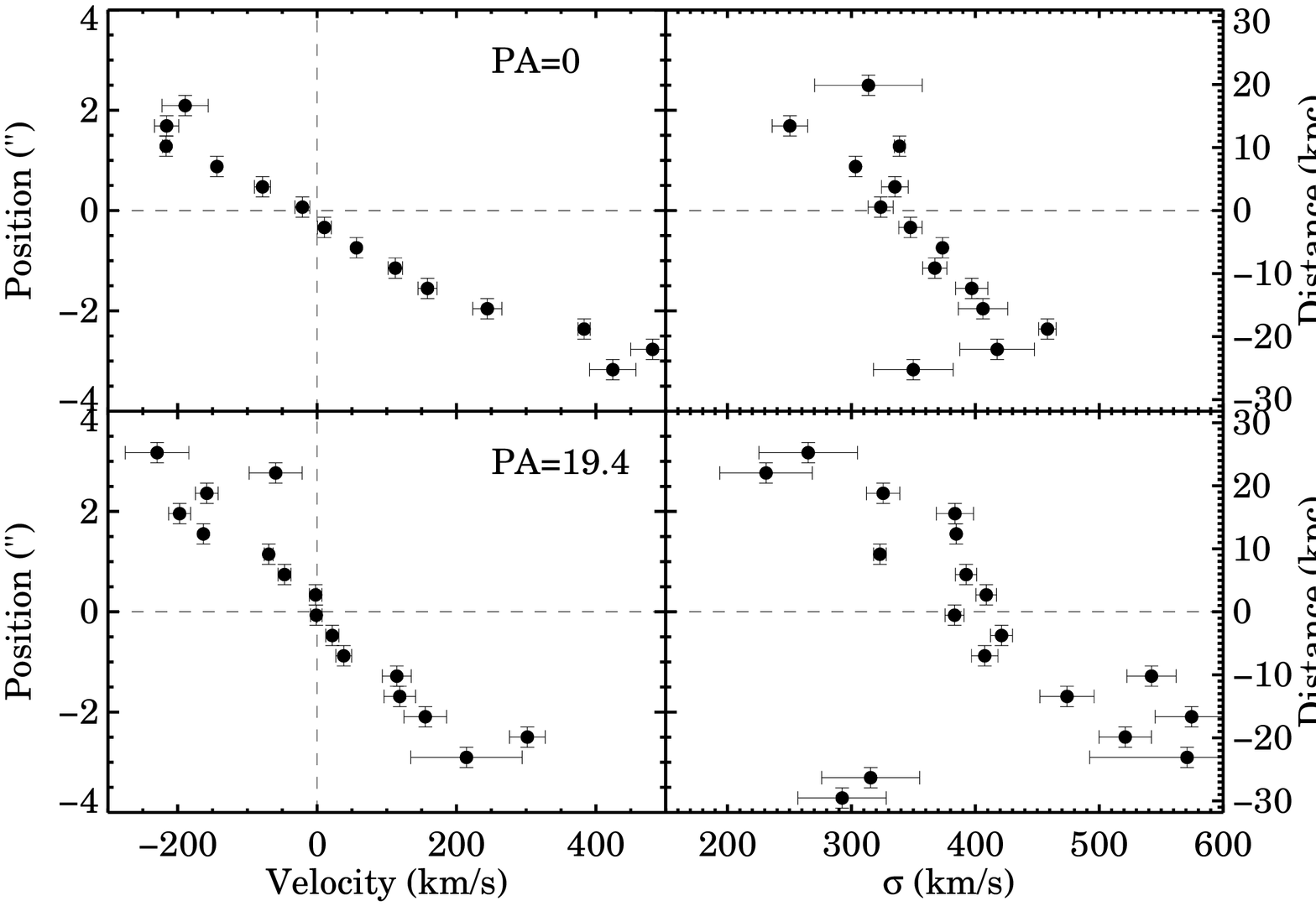}
\caption{Kinematics of the gas in the \lya\ nebula surrounding \blob. The top two panels show the spatial variation in central velocity (left) and velocity dispersion (right) determined from Gaussian fits to the \lya\ emission line profile measured using the PA=0$^\circ$ observation. The lower panels show the measurements from the PA=19.4$^\circ$ observation. The origin for the ordinate is defined as the spatial location of the centroid of the \lya\ emission (i.e., the same as in figure~\ref{specslits}). The velocity origin is defined using the wavelength of the \ion{He}{2}$\lambda$1640 emission line.}
\label{velblob}
\end{center}
\end{figure*}

Figure~\ref{velblob} shows the velocity structure in the \lya\ line as determined by these fits. The seeing for both observations was typically twice the aperture, and neighboring points in the figure are therefore correlated. The nebula exhibits a monotonic change across the central $\approx$4\arcsec\ in both the central velocity and the velocity dispersion, with deviations from this trend only near the extrema where the signal-to-noise ratio is low.

The remarkable aspect of the measured profile is that its velocity varies so uniformly across the central region. Fitting a straight line to the data from the two observations using a robust least absolute deviation algorithm results in a velocity profile across this central region of $v = 36.2 - 124.1r\ {\rm km\ s^{-1}}$ in PA=0$^\circ$ and $v = 9.7 - 75.5r\ {\rm km\ s^{-1}}$ in PA=19.4$^\circ$, where $r$ is the angular separation in arcseconds. This smooth variation is unlikely to be due to a smooth change in the absorption by neutral hydrogen, since the line profile does not show strong evidence for this. If it is due to the gas kinematics, then the variation could be due to infall, outflow or rotation, and it is not easy to discriminate between these options with the present data.

\subsection{Spectral Energy Distribution}

\begin{figure*}
\begin{center}
\plotone{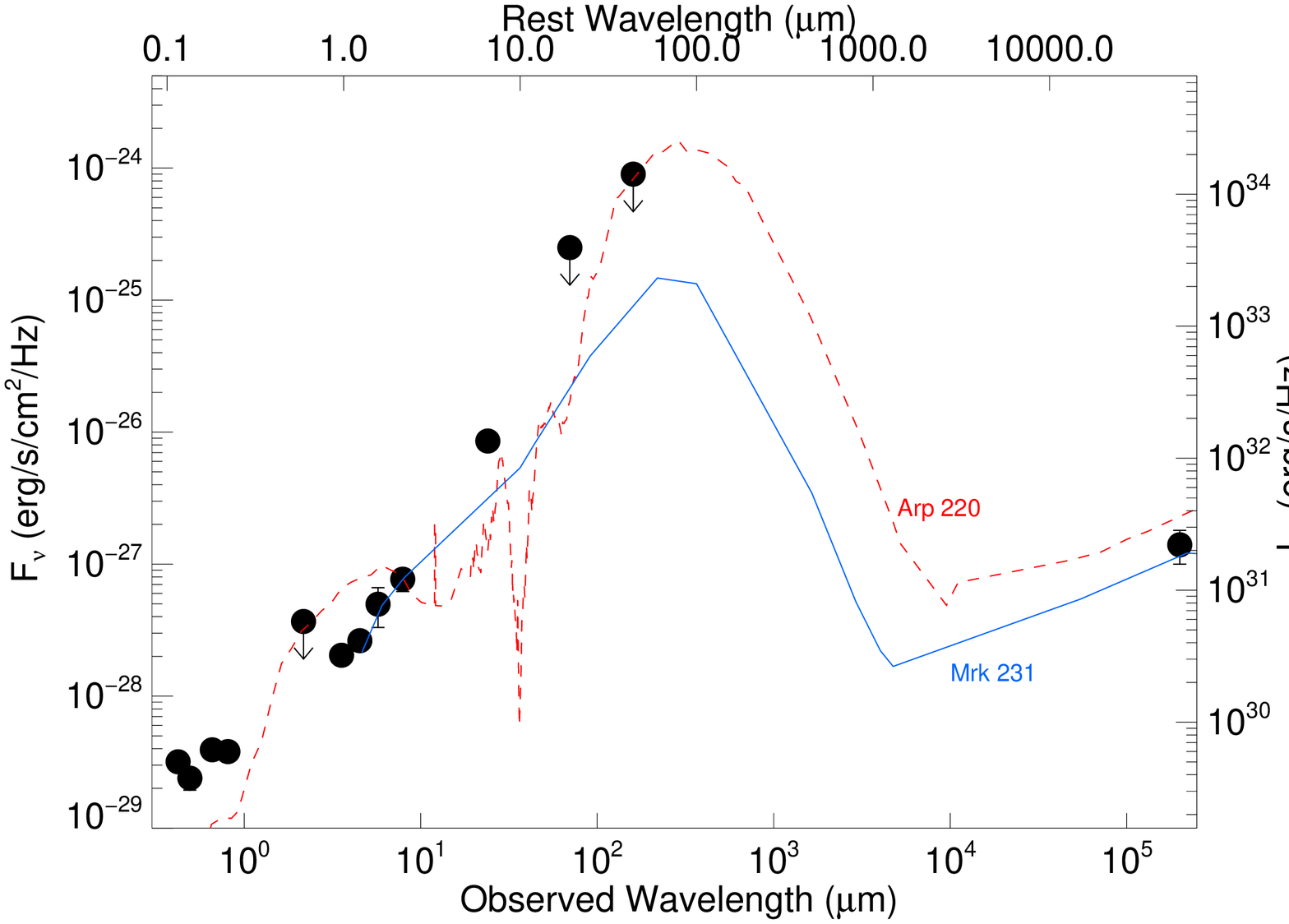}
\caption{The spectral energy distribution of \blob. The optical points ($\lambda_{\rm obs} < 1\micron$) should be regarded as upper limits, since they include continuum contributions from galaxies A and B. The red dashed curve is the SED of Arp~220 ($\lambda < 5\micron$ from \citet[]{spo2004a} and $\lambda > 5\micron$ from \citet[]{sil1998}), scaled in luminosity by a factor of 26, for comparison. Note the discrepancy between the two objects near $\lambda_{\rm obs}\approx10-20\micron$, where the SED of Arp~220 transitions from being dominated by starlight to being dominated by dust emission; \blob\ shows no obvious evidence for such a transition. The blue solid line represents the SED of Mrk~231 from \citet[]{ivi2004}, scaled by a factor of 2.}
\label{sed}
\end{center}
\end{figure*}

Figure~\ref{sed} shows the spectral energy distribution (hereinafter SED) of \blob. The optical data points should be regarded as upper limits, since they include continuum contributions from galaxies A and B. 

A model for the SED of Arp~220 (constructed from the models of \citet[]{sil1998} and \citet[]{spo2004a}) placed at $z=2.6562$ and scaled by a factor of 26 is shown for comparison in figure~\ref{sed}. The scaling factor was chosen to match the observed IRAC 8\micron\ flux density of \blob. The SED of Arp~220 differs from that of \blob\ in two ways. First, the rest-frame UV emission observed in \blob\ is much larger than that from Arp~220. This is not surprising: as mentioned before, the UV points are contaminated by other sources (and the bluest point by the strong \lya\ line emission). Second, the 5.8\micron\ and 8.0\micron\ IRAC points lie significantly above the prediction of the Arp~220 template. This region of the spectrum is of interest because it is where the SED of Arp~220 tranisitions from being dominated by starlight at shorter wavelengths to being dominated by dust emission at longer wavelengths. The SED of \blob\ shows no evidence for such a transition, suggesting that perhaps there is little or no significant contribution from starlight in this portion of the spectrum. Indeed, the mid-infrared continuum measurements from Spitzer ($3.6\micron - 24\micron$) are well-modeled by a simple quadratic power-law, $f_\nu\propto\nu^{-2}$, also suggesting that the Spitzer bands are not strongly affected by emission from polycyclic aromatic hydrocarbons (PAHs), of which the strongest features are redshifted out of the IRAC bandpasses.

The mid-infrared color of \blob\ lies in the region of parameter space more generally occupied by quasars than by normal star-forming galaxies \citep[e.g.,][]{ste2005,lac2004}. Figure~\ref{sed} also shows the SED of the Seyfert 1 galaxy Mrk~231 \citep[e.g.,][]{ivi2004} placed at the redshift of \blob\ and scaled by a factor of 2. The SED of Mrk~231 is a better fit to the observed 3.6\micron\ to 24\micron\ photometry of \blob, and suggests that this source is an enshrouded AGN. The observed excess in the MIPS 24\micron\ band over the Mrk~231 SED may be due to 7.7\micron\ PAH emission.

The most likely emission mechanism responsible for the mid-infrared luminosity is dust emission. Emission at these wavelengths would require dust at warm temperatures ($\gtsim 500$ K), higher than those typically seen in normal or ultraluminous star-forming galaxies. However, a power-law spectrum of $F_\nu\propto\nu^{-2}$ can also be produced by synchrotron radiation from power-law distribution of electrons, with $N(E)\propto E^{-5}$. At longer wavelengths, synchrotron self-absorption sets in and the spectrum should take on a $\nu^{2.5}$ shape. We can use the faint radio detection of 0.14~mJy and estimate the turnover frequency to be $\sim 1.2\times 10^{11}$~Hz (2.5mm). This is not a very useful constraint, especially since a synchrotron spectrum of this sort has never been observed in an extragalactic source. If the slope of the radio spectrum is similar to that of Mrk~231 and its extrapolation to the mid-infrared significantly undershoots the observed  mid-infrared flux densities, then the emission mechanism for the mid-infrared is thermal emission from dust grains. Observations in the sub-mm to radio wavelengths will be necessary to better constrain the radiation mechanism for the IR portion of the SED. 

If \blob\ has a similar SED to that of Arp~220 (Mrk~231), the comparison suggests that it has a total far-infrared luminosity of $L_{\rm IR}\ltsim 7.5\times 10^{13}\Lsun$ ($1.1\times 10^{13}\Lsun$)! We emphasize, however, that these estimates are very uncertain at present, since the bulk of the luminosity emerges at wavelengths at which the flux density of \blob\ is currently unknown and require a large bolometric correction. Measurements at sub-mm wavelengths are clearly crucial to establishing the shape of the SED near its peak and providing a better measure of the bolometric luminosity. If the SED is similar to that of Arp~220 (Mrk~231), we would expect 450\micron\ and 850\micron\ flux densities of $\ltsim$120~mJy (8~mJy) and $\ltsim$40~mJy (2~mJy) respectively. Clearly, these predictions are all within the reach of the current sub-mm observations. 

We note that the mid-infrared-to-radio flux density ratio observed in \blob, $q_{24}\equiv {\rm log}[f_\nu(24\micron)/f_\nu(20cm)] \approx 0.8$, shows remarkable similarity to that of starburst galaxies, which typically have $q_{24}\equiv {\rm log}[f_\nu(24\micron)/f_\nu(20cm)] = 0.84\pm0.28$ \citep[]{app2004}. Mrk~231 also falls on this relation, although it is an obscured Seyfert 1; this is because its mid-infrared and radio emission are dominated by flux from the circumnuclear starburst rather than the active nucleus. 

\section{Discussion}

In the previous section, we presented our various observations of \blob. Here, we discuss the implications of our observations for understanding the physical nature of this remarkable system.

Large \lya\ nebulae such as \blob\ are rare. Thus far, most known large \lya\ nebulae are associated with powerful radio galaxies \citep[e.g.,][]{mcc1990, van1997,pen1997,dey1997,dey1999mdrg,vil2002,vil2003,reu2003}. Among radio quiet sources, there are two similarly large nebulae at $z\approx 3.09$ \citep[\lya\ ``blobs'' LAB1 and LAB2;][]{ste2000} and one at $z\approx 2.38$ \citep[2142$-$4420~B1;][]{fra1996}. These latter systems are associated with large galaxy overdensities, and follow-up deeper \lya\ imaging of these regions has resulted in the discovery of numerous other smaller \lya-emitting nebulae \citep[e.g.,][]{pal2004,mat2004}. Similar overdensities have also been found associated with some of the \lya\ nebulae associated with radio sources \citep[e.g.,][]{kee1999,pen2000}.

The rarity of these $>$100~kpc \lya\ clouds, their association with powerful AGN and galaxy overdensities, and their energetics, all suggest that these regions are the formation sites of the most massive galaxies. If so, understanding the physical conditions and energetics of these systems can provide important insights into the massive galaxy formation process. 

At present, it is still not clear what powers the nebulae. Is the \lya\ emission cooling radiation of infalling, shocked, pristine gas \citep[]{hai2000}? Or is it photoionized by hot stars or AGN \citep[e.g.,][]{hai2001}? Or is it the cooling of outflowing shocked gas \citep[cf. the ``superwind model'' of ][]{tan2000,fra2001}? Much of the recent literature on the extended \lya\ nebulae has focussed on the radio-quiet systems (LAB1, LAB2 and the sources at $z=2.38$), perhaps because the absence of an obvious radio source suggested an absence of AGN (and their associated complexity) in these systems. However, more recent observations of the LAB1 system now find that it may also contain buried AGN \citep[]{cha2004a}. Moreover, there are many similarities between the outer, kinematically quieter regions of the \lya\ halos surrounding the powerful radio sources and the radio quiet ``blobs''. All this suggests that AGN may exist in all these nebulae: after all, if all massive bulges host massive black holes, then perhaps the formation of the active nucleus accompanies the formation of the galaxy. 

It is then perhaps not surprising that the known large \lya\ nebulae reveal such complexity:  compact obscured regions, formed galaxies, revealed AGN, all within their extent. If these are the largest protogalactic regions, one might expect that both star- and AGN-formation could proceed within the cloud. 

\subsection{Mass of the Nebula}

The total mass of the nebula can be estimated dynamically (from the velocity profile of the \lya\ line), under the relatively poor assumptions that the line profile is not significantly affected by absorption and scattering. The smooth variation of the central wavelength of the line, the relatively constant velocity dispersion, and the overall symmetry of the line all provide some justification to these assumptions. If the velocity profile observed in the inner regions is primarily the result of rotation, then the mass internal to a radius $r=4\arcsec$ ($\approx 32~kpc$) is $M(r\le4\arcsec) \approx 0.6-1.6\times10^{12} \ {\rm sin^{-2}}i~\Msun$, where the large range is due to the range in the velocity profile fits measured from the two different spectroscopic observations. The velocity profile is not well measured at radii larger than $\sim 30$~kpc.

Interpreting the line width is also fraught with uncertainty. Absorption by neutral hydrogen would result in the measured width being an underestimate of the true width; whereas resonant scattering would likely broaden the line. The velocity dispersion of $\approx400\ {\rm km\ s^{-1}}$ is too large to be due to the thermal motions of the gas.  If it is, instead, due to the random motions of clouds within the nebula, then we can estimate the dynamical mass of the nebula to be $M(r\le4\arcsec) \approx 6\times10^{12}~\Msun$. These dynamical mass estimates derived from the line centroid and width are very large. However, they are surprisingly similar to the estimates derived for the large \lya\ halos around high-redshift radio galaxies, where dynamical mass estimates range from $0.3 - 10\times 10^{12}~\Msun$ \citep[]{reu2003,vil2003,van1997}. We note that in general such dynamical estimates are more sensitive to larger masses, since lower mass systems will have gradients that are not resolved by typical faint-object spectroscopic observations.

A separate estimate of the ionized gas mass may be derived by estimating the electron density $n_e$ from the net luminosity of the \lya\ emission line \citep[cf.][]{mcc1990}, $M_{\rm ion} = 1.25m_pn_ef_VV\approx 1.7\times 10^{12}f_V^{0.5} (n_ef_V^{0.5}/0.026{\rm cm^{-3}}) \Msun$. The filling factor $f_V$ is unconstrained by the present data, but observations of line-emitting regions in cluster cooling flows suggest filling factors of order $10^{-5}$ to $10^{-6}$ \citep[e.g.,][]{hec1989}, which would suggest ionized gas masses of $M_{\rm ion}\approx 1.7\times 10^{9} (f_V/10^{-6}) \Msun$. If $\Omega_b=0.04$ and $\Omega_{\rm matter}=0.30$, then a comparison with the dynamical mass estimates suggest ionization fractions of $\ltsim 0.013 (f_V/10^{-6})^{0.5}(10^{12}/M_{tot})$, and a rough upper limit to the filling factor of $\sim 7.7\times 10^{-5}$. The ionized gas fraction is low for a typical hot plasma, suggesting that a significant fraction of the mass might be contained in colder (neutral), collapsed objects, e.g., the galaxies that are forming within the cloud. Although this argument is only speculative at present, better measures of the gas density, temperature and filling factor, perhaps from deep rest-frame optical spectroscopy, can help constrain the relative fractions of hot and cold gas.

\subsection{Physical State of the Gas and the Nature of the Ionizing Source}

It is difficult to place firm constraints on the temperature of the gas cloud since at present \lya\ is the only emission line that we detect across the spatial extent of the nebula; \ion{He}{2}$\lambda$1640, \ion{C}{4}$\lambda\lambda$1550 and \ion{C}{3}]$\lambda$1909 are only detected in a localized region near the center of the nebula, and in addition, \ion{C}{4} and \ion{C}{3}] are observed in a red, optically visible continuum source which may be associated with the mid-infrared source. There is also weak evidence that \ion{C}{4} may be spatially extended between the central region and the MIPS source. 

The line ratios of the gas in the central region are not representative of shock ionization: the shock models of \citet[]{dop1996} typically produce much higher \ion{C}{4}$\lambda\lambda$1550/\ion{He}{2}$\lambda$1640 ratios (e.g., $\sim 2-20$ for shock velocities $v_{\rm shock}\sim 500-150\ {\rm km\ s^{-1}}$) as compared to the observed value of \ion{C}{4}/\ion{He}{2} and tend to overproduce NV$\lambda$1240 relative to what is observed in this source. Higher shock velocities ($v_{\rm shock}>500\ {\rm km\ s^{-1}}$), where better matches might be produced, are unlikely given the low velocity widths observed for the \ion{C}{4} and \ion{He}{2} emission lines. We cannot yet, however, rule out shock ionization in other regions of the nebula where \lya\ is the only detected emission line.

The \ion{He}{2}$\lambda$1640 emission line is unusually strong in the central region. Its strength may be indicative of low metallicity (and therefore high temperatures) in the ionized gas in this region. However, models of star-forming galaxies typically only show \ion{He}{2} with rest-frame equivalent widths of $\gtsim 15$\AA\ at very low metallicities ($Z\ltsim 10^{-7}$) and at very young ages $\ltsim 2$~Myr \citep[]{sch2003}. Although it is possible we are witnessing this object at such an unusual and early epoch, it would be very improbable; it is therefore unlikely that the strong \ion{He}{2} is due to low metallicity gas.  

Harder ionizing continua, such as those generated by AGN, can also produce strong \ion{He}{2}; indeed, comparable strengths of \ion{C}{4} and \ion{He}{2} are commonly observed in radio galaxy spectra and ``type 2 quasars'' and are generally attributed to photoionization by a power-law continuum. The one oddity is the relative weakness of the observed \ion{C}{3}] emission: most radio galaxy spectra show \ion{C}{3}]/\ion{C}{4} $\sim$ 1, whereas we observe a ratio in this system of $\approx 0.2$. The observed ratio can however be produced by a power-law ionizing spectrum and is characteristic of a large ionization parameter $U\equiv Q({\rm H})/4\pi R^2n_ec\sim10^{-2}$ for a $f_\nu\propto \nu^{-1}$ continuum \citep[]{fer1981} in this central region. 

A crude constraint on the slope of the ionizing spectrum can be derived from the ratio $\eta$ of the number of He$^+$ ionizing photons ($\approx 3.7\times 10^{53}{\rm \ s^{-1}}$) relative to the number of H ionizing photons ($\approx 1.6\times 10^{55}{\rm \ s^{-1}}$):  $\eta\equiv Q({\rm He}^+) / Q({\rm H}) = 0.023$ (for a $T=10^4$ K gas) \citep[e.g.,][]{sch2002}. If we relate this ratio to the number of photons at 54 eV relative to that at 13.6 eV, this ratio suggests a spectral shape of $F_\nu \propto \nu^{-2.7}$. It is intriguing that this slope is even steeper than the $F_\nu\propto\nu^{-2}$ spectrum derived by extrapolation of the mid-infrared SED. We have not accounted for \lya\ absorption in this estimate, which would suggest that the ionizing spectrum is even steeper, i.e., $\alpha < -2.7$. Note that here we have compared the full 10\arcsec\ radius aperture measurement of \lya\ with the spectroscopic measurement of \ion{He}{2}$\lambda$1640. If we compare just the measurements in the same spectroscopic apertures, we derive a ratio of $\eta = 0.21$ for $T=10^4$~K, which would suggest an extremely hard ionizing spectrum of $F_\nu\propto\nu^{-0.116}$. This latter estimate is unlikely to be realistic, since the H$^+$ ionization zone is likely to extend over a much larger scale than the He$^{++}$ zone, but we can take it to be a very firm upper limit for the slope of the ionizing continuum. 

If we assume normal HII region electron temperatures of $T_e\sim 10^4$~K in the central region, then we can estimate the electron density using the strength of the \lya\ emission in the same aperture used to measure the high ionization lines (see Table~\ref{specdata}). The \lya\ luminosity is given as $L_{\rm Ly\alpha} = (j_{\rm Ly\alpha} / j_{\rm H\beta})n_en_pf_VVh\nu_{\rm H\beta}\alpha_{\rm H\beta}^{eff} \approx 4.05\times10^{-24}n_e^2f_VV$, where $V$ is the volume of the emitting region, $n_e$ and $n_p$ are the electron and proton number densities, $f_V$ is the filling factor of the ionized gas, and $\alpha_{\rm H\beta}^{eff}$ is the effective recombination coefficient for H$\beta$ \citep[]{sto1995}. In the central region, we find $n_e f_V^{0.5}\approx0.25\ {\rm cm^{-3}}$; combined with the ionization parameter estimate, this implies a hydrogen ionizing photon flux in the central region of $Q({\rm H})\approx 3\times10^{54}f_V^{-0.5}\ {\rm s^{-1}}$ photons s$^{-1}$. We have no direct measure of the electron density, and therefore no firm constraints on the volume filling factor of the ionized gas. However, if we use the rough upper limit to the filling factor from the previous subsection, then $Q({\rm H}) \gtsim 6\times 10^{56}$ photons s$^{-1}$.

Using a similar analysis for the total \lya\ luminosity measured in a 10\arcsec\ radius aperture, we find $n_ef_V^{0.5}\approx0.026\ {\rm cm^{-3}}$ for the average density in the nebula as a whole. For values of the filling factor $10^{-3}>f_V>10^{-6}$, the average electron density in the line-emitting regions is in the range $0.8 < n_e < 26\ {\rm cm^{-3}}$.

\subsection{Energetics}

The \lya\ luminosity of the nebula within a 10\arcsec\ radius aperture is $\approx 1.67\times10^{44}\ {\rm erg\ s^{-1}}$. This is comparable to the luminosities of the other well-known \lya\ nebulosities, LAB1 \citep[$1.4\times10^{44}$]{ste2000} and 2142$-$4420~B1 \citep[$\approx 8\times10^{43}\ {\rm erg\ s^{-1}}$]{fra2001,pal2004}. If it is due to young, hot stars distributed over the face of the nebula in a very extended star-forming region, then the luminosity implies a net star-formation rate $SFR_{\rm Ly\alpha} = L_{\rm Ly\alpha}/(1.12\times10^{42}\ {\rm erg\ s^{-1}}) \approx 150~\Msun\ {\rm yr^{-1}}$ \citep[]{ken1983}. 

It is not clear what source(s) power the \lya\ emission. No obvious single compact source with high UV luminosity is visible within the extent of the nebula. The two galaxies A and B do not appear to be plausible candidates for the sources of the photoionizing radiation: galaxy B is at the wrong redshift, and galaxy A does not show any strong \lya\ emission associated with it and its UV luminosity at $\lambda_{\rm rest} = 1500$\AA\ only corresponds to a modest star-formation rate of $SFR_{\rm UV} = L_{\rm UV}/8\times10^{27}\ {\rm erg\ s^{-1}\ Hz^{-1}}\approx 8\ \Msun\ {\rm yr^{-1}}$ \citep[e.g.,][]{mad1998}. 

It is also not clear if the MIPS source is the source of the photoionizing radiation: it is faint at rest-frame UV wavelengths and its mid-infrared emission suggests that the source, although very luminous, is almost completely obscured from view by dust. It is therefore unlikely that short wavelength ionizing photons could escape from the nebula - clearly, little or no \lya\ emission does. Nevertheless, it is intriguing that the estimated far-infrared luminosity of the MIPS source is roughly two and a half orders of magnitude larger than the \lya\ luminosity of the entire nebula, and it is possible that some ionizing radiation could leak out in directions other than our direct line of sight.  We discuss this further below. 

It therefore appears that the \lya\ emission is powered by sources hidden from direct view, perhaps partly by the mid-infrared source, perhaps partly by another obscured or mis-directed source lying near the center of the nebula where we also observe the \ion{C}{4} and \ion{He}{2} emission lines, or by distributed, faint, ionizing sources, or by shocks associated with the infall of gas into the region. The dynamical timescale of the nebula is given by $t_{\rm dyn}\approx {\pi\over{2}}\sqrt{\rm R^3/2GM} \approx 0.37\ {\rm Gyr}$. This is much smaller than the age of the Universe at $z=2.656$ (2.4~Gyr in our adopted cosmology), and suggests that we are witnessing the nebula in a young phase. The temperature of the nebular gas is unknown, but if we assume a temperature of $10^4$~K, then the cooling timescale is $t_{\rm cool} \approx {{3\rho k T}\over{2\mu \Lambda(T)}} \approx  2.07\times10^{11} s (T/10^4{\rm K})(n_e/1{\rm cm^{-3}})^{-1}(\Lambda(T)/10^{-23}{\rm erg\ cm^3\ s^{-1}})^{-1}$ \citep[]{pad2002,sut1993}. For gas between $10^4-10^5$, log($\Lambda(T)$) can vary from $-23.4$ to $-21.0$ depending on the exact temperature and metallicity, suggesting cooling times which are very short ($t_{\rm cool} \ltsim 1.6\times10^4{\rm year}$), implying that some source of energy (shocks? ionizing photons?) is required to maintain the nebula and render it visible. Taken at face value, the large ratio of $t_{\rm dyn}/t_{\rm cool}$ would imply rapid collapse of the cloud into star-forming clumps.

\subsection{The Mid-Infrared Source}

The source \blob\ detected in the {\it Spitzer} IRAC and MIPS bands lies within the extent of the \lya\ nebula, roughly 2\farcs5 (20~kpc) away from the center. The optical spectra show a faint continuum source within the MIPS error circle, just to the southeast of the MIPS position. The spectrum of this source appears to be significantly redder than galaxy A ($F_\lambda \propto \lambda^{0.075\pm0.006}$ for the MIPS source in the observed optical regime, versus $F_\lambda \propto \lambda^{-0.076\pm0.004}$ for Galaxy A), but it does show weak associated \ion{C}{3}] emission and possibly some \ion{C}{4} emission certifying its association with the \lya\ nebula. The optical spectrum shows no obvious signs of AGN activity (e.g., broad emission lines or strong high-ionization lines), but instead shows possible absorption lines similar to those observed in young, star-forming Lyman-break galaxies. 

\begin{figure*}
\begin{center}
\plotone{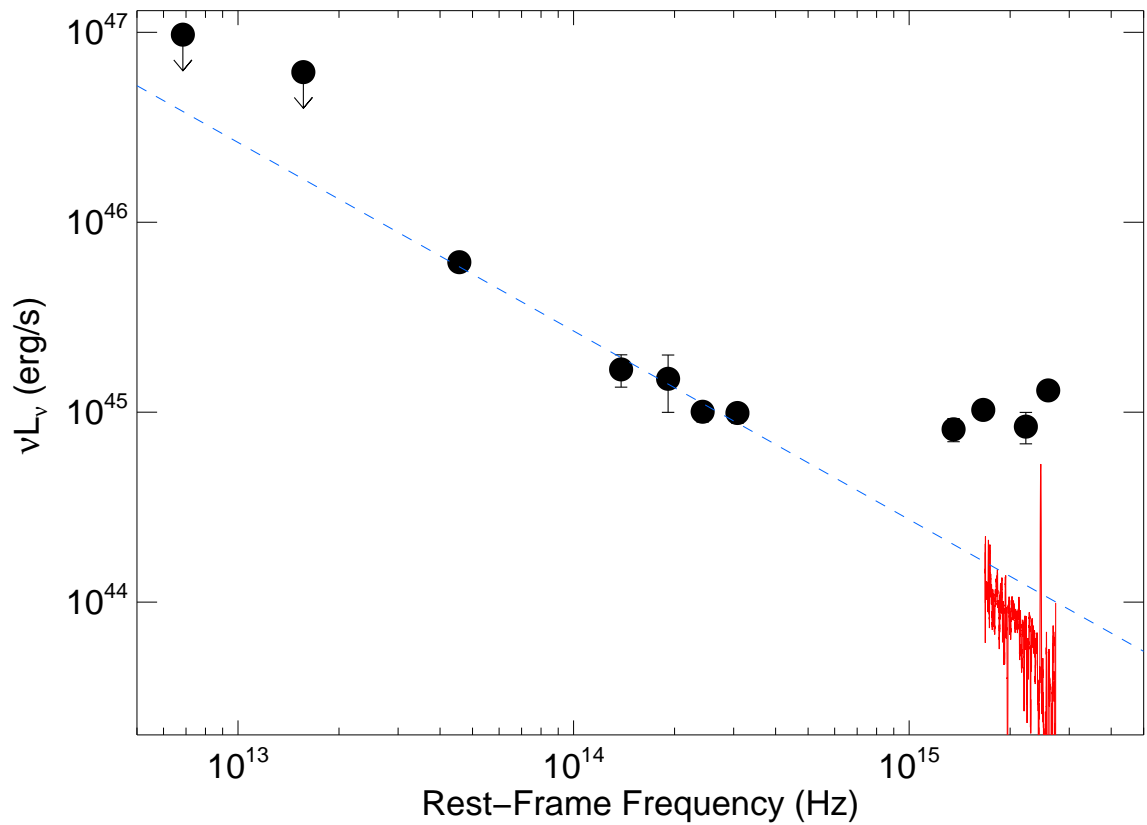}
\caption{The spectral energy distribution of \blob, shown here as $\nu L_\nu$ {\it vs} $\nu$. The rest-frame UV points ($\nu > 10^{15}$ Hz) should be regarded as upper limits, since they include continuum contributions from galaxies A and B. The red line at high frequency is the spectrum of the optical continuum source closest to the location of the MIPS source (shown also in figure~\ref{spec1d_mips}). The dashed blue line is a fit to the {\it Spitzer} data points, $L_\nu \propto \nu^{-2}$. }
\label{sed2}
\end{center}
\end{figure*}

A fit to the SED in the {\it Spitzer} bands suggests a power-law spectrum of the form $L_\nu \approx 2.7\times10^{31}(\nu/10^{14}{\rm Hz})^{-2}{\rm \ erg\ s^{-1}\ Hz^{-1}}$ (figure~\ref{sed2}). Rather surprisingly, extrapolating this power-law to rest-frame UV wavelengths 1300$-$1800\AA\ is consistent with the observed spectrum of the faint optical source believed to be associated with the MIPS source, only overshooting the observed spectrum by a factor of $\sim$1.3$-$1.7 (figure~\ref{sed2}). This offset is perhaps not surprising, given that we are comparing the large aperture total flux densities of the {\it Spitzer} data with the small aperture (1.5\arcsec $\times$1.0\arcsec) spectral extraction. 

The mid-infrared source has poorly constrained flux densities at long wavelengths ($\gtsim 100$ \micron) where much of its luminosity likely emerges. As discussed in section \S~3.6, if the SED is similar to that of known ULIRGs, the far-infrared luminosity of the source is likely to be in the range $1-5 \times 10^{13}\Lsun$. If this object is a starburst, this luminoisty corresponds to a star-formation rate $SFR_{\rm IR}\approx 1700(L_{\rm IR}/10^{13}\Lsun){\rm \Msun\ yr^{-1}}$. If its luminosity-to-mass ratio is comparable to ULIRGs (i.e., $L/M\sim 100-200$), the prodigious luminosity implies a total mass in dust of about $M_{\rm dust}\sim 0.2-5\times 10^9\Msun$ if $M_{\rm dust}/M_{\rm gas}=0.01$. These estimates are very coarse, but it is clear that these large bolometric luminosities imply very large masses of dust, and that a substantial mass of baryons has already been cycled through stars to produce this mass of dust. 

Although the mid-infrared source is extremely energetic and lies within the nebula, it is not clear whether it is responsible for photoionizing the nebula. If we assume that the $\nu^{-2}$ power-law represents the intrinsic spectrum of the mid-infrared source, and if it can be extrapolated to shorter wavelengths, then we can investigate whether the ionizing photons emitted by this source are sufficient to photoionize the nebula and produce the observed \lya\ radiation. Extrapolating the power-law would result in $\approx 1.8\times10^{54}$ ionizing photons between 200\AA\ and 912\AA. The luminosity of the \lya\ nebula requires $\approx 10^{55}$ ionizing photons. The MIPS source can therefore be responsible for at most 18\% of the \lya\ luminosity. This is likely a very strong upper limit, since in reality the rest-frame UV spectrum is observed to be redder than the extrapolation, and our estimate here does not account for any absorption by dust. It is possible, of course, that although the extinction to the source is severe along our line of sight, it may still have low optical depth toward the nebula (say, a mis-directed quasar). Although plausible, the geometry of the source, taken at face value, does not favor such an explanation:  the MIPS source is not located at the center of the nebula and instead lies $\approx$20~kpc away from the peak of the \lya\ emission. The symmetry of the \lya\ emission around this peak, the presence of a possible ionizing source near this peak as witnessed by the presence of \ion{He}{2}, \ion{C}{4} and \ion{C}{3}] emission, and the lack of strong \lya\ emission cospatial with the location of the MIPS source all suggest that it is not primarily responsible for the nebula. Further spectroscopy and high spatial resolution imaging of the cloud are necessary to carefully account for all the possible ionizing sources within the cloud.

The power-law SED in the optical and mid-infrared bands exhibited by \blob\ and its extreme luminosity are more typical of active galaxies and quasars than of star-forming galaxies \citep[e.g.,]{pie1992,nen2002}. It is therefore likely that the infrared luminosity of this source is dominated by an enshrouded AGN, rather than a vigorous starburst; starlight from young stars only begins to dominate the flux at rest-frame UV wavelengths. In this respect, \blob\ may be very similar to the dust enshrouded systems with extreme 24\micron-to-optical colors discovered by \citet[]{hou2005}. {\it Spitzer} IRS spectroscopy of these extreme systems has established that they are obscured, dust-enshrouded AGN at $z\sim 2.5$ with far-infrared luminosities in the range $6\times 10^{12-13}~\Lsun$ \citep[]{hou2005}.  We speculate that these objects are all extremely massive galaxies caught in an early phase of their formation. Since the AGN is currently dominating the SED, it is tempting to speculate that the supermassive black hole in these extremely massive systems form contemporaneously with the assembly of the galaxy, rather than grow slowly by subsequent accretion.  

\subsection{Galaxy and AGN Formation in the Cloud}

In a seminal work, \citet[]{elm1977} proposed that star formation in molecular clouds in our Galaxy proceeds through the sequential formation of OB subgroups, with each new subgroup driving ionization and shock fronts deeper into the parent molecular cloud, stimulating further star formation. Their model was inspired by observations of a few Galactic star-forming regions (e.g., W3, M~17 and M~42), where  this sequence of events can be observed directly: an expanding HII region associated with an OB association located at the edge of a molecular cloud. A host of infrared-luminous cores and maser sources lying within the molecular cloud and near this edge bear witness to active, ongoing star-formation occurring near the location of the shock front, and suggest that the star-formation ``front'' is slowly eating into the cloud. 

It is rather remarkable that the system reported here shows many analogous features: a Lyman break star-forming galaxy, which lies just north of a dust-enshrouded active region, which lies just north of a large region characterized by \lya\ cooling radiation. By analogy to the M~17 region, we postulate that this is a time sequence, with the Lyman break galaxy representing the `old', revealed objects, the mid-infrared source representing the site of current star- and AGN-formation, and the central source and \lya\ nebula representing the region where galaxy formation is just beginning. In such an evolutionary scenario, a gas cloud, observed by its cooling radiation, would undergo collapse leading to an initial burst of star-formation and AGN activity. This immediately enshrouds the active region (through the production of dust in the ejecta of  first generation supernovae), and then finally leads to a stage when the formed, star-forming galaxy is unveiled. Although highly speculative, this scenario would certainly have many merits. It might explain the low stellar masses observed in the \lya-bright regions found in the narrow-band surveys (these are the youngest objects that have just started forming stars), the very high star-formation rates observed in the SCUBA galaxies (these represent the peak star-formation phase), the relatively low star-formation rates seen in the Lyman-break galaxies (the early post-starburst phase), and no obvious evidence of systems with intermediate metallicities (they are rapidly enshrouded). 

In this case, however, it is very unlikely that the formation events could be triggered by an ionization front: over the several kpc scales in question, the only relevant force is gravity, and the hypothesis presented here is simply that we may be able to estimate the time for the collapse of such a large cloud into its star-forming sub-units by age-dating the various sub-units. If, for example, galaxy A is typical of most Lyman-break galaxies at $2<z<3$, then we can assume a minimum age of $\sim 100$~Myr since its last major epoch of star-formation \citep[e.g.,][]{pap2001}. Since its distance from the center of the \lya\ emission is $\approx 30$~kpc, a naive estimate of the rate at which galaxy formation is proceeding through the region is $\sim 300\ {\rm kpc\ Gyr^{-1}}$. Better constraints may be derived by detailed age-dating of the galaxies in the nebula. Our present data preclude such an analysis, but future spectroscopic and multiwavelength imaging observations may allow us to date the various components more reliably. 

One timescale of relevance here is the formation time of the AGN associated with the mid-infrared source.  Under the assumption that the bulk of the far-IR luminosity is produced by accretion onto a super-massive black hole (SMBH) at the Eddington rate, we estimate a central black hole mass of $M_{\rm BH} \approx 2.5\times 10^8 (L_{IR}/10^{46}{\rm erg\ s^{-1}})\Msun$ and an accretion rate of $\Mdot \approx 7.6 (L_{IR}/10^{46}{\rm erg\ s^{-1}})(0.1/\eta)\Msun\ yr^{-1}$, where $\eta$ is the radiative efficiency of the SMBH. If the black hole grows at a constant efficiency, then a $10^9 \Msun$ SMBH can be created by steady accretion from a small seed $10 \Msun$ black hole in about $8 \times 10^8$ years, i.e., a seed formation redshift of $z\approx4$.  Our estimates are upper limits for $M_{\rm BH}$ and the total growth time.  For more massive seeds or merging of multiple seeds during the formation process, the formation time could be significantly shorter \citep[cf.][]{hai2004,yoo2004,sha2005}.

\section{Conclusions}

We have discovered a very large \lya-emitting nebula at a redshift of $z=2.656$. This is the first such object discovered by its associated 24\micron\ emission. The luminosity of the nebula ($L_{\rm Ly\alpha} \approx 1.67\times 10^{44}\ {\rm erg\ s^{-1}}$ within a 10\arcsec\ radius aperture), its large spatial extent ($>100~{\rm kpc}$), the detection of a star-forming ``Lyman-break" galaxy and a source of intense mid-infrared emission within the nebula all suggest that it is the site of massive galaxy formation. In many of these respects, \blob\ is similar to the \lya\ ``blob'' LAB1 discovered by \cite{ste2000}. By analogy to the other large \lya\ ``blobs'' known, it is probable that this one sits within a large (several Mpc - scale) structure; further deep, wide-field \lya\ imaging of the region should prove fruitful. 

Spectroscopy of the \lya\ emission reveals that the cloud is kinematically `quiet' at least in its inner regions, showing a slowly and monotonically varying velocity centroid with position and roughly uniform velocity dispersion across its face. If interpreted as rotation, the implied dynamical mass contained within the central 8\arcsec\  is $\sim 0.6-1.6\times 10^{12}~\Msun$. The velocity dispersion of the \lya\ emission line is large ($\sigma \ltsim 400\ {\rm km\ s^{-1}}$). 

The  source of the ionization for the nebula is ambiguous. The geometry of the emitting region and the shape of the SED suggest that the mid-infrared source \blob\ is unlikely to be responsible for more than $\sim$20\% of the \lya\ luminosity, and it is likely that the ionization and line radiation are the result of shocks and / or multiple, distributed ionizing sources. If the \lya\ emission is primarily powered by hot, young stars, the total star formation rate of the system is $\sim 150~\Msun\ {\rm yr^{-1}}$. The detection of localized, weak, narrow \ion{C}{4} and \ion{He}{2} emission near the center of the nebula suggests that at least portions of the nebula may be ionized by a power-law ionizing source or perhaps contain low metallicity gas; however, there is clearly 24\micron\ emission from other parts of the system that most likely is due to emission from dust, suggesting that the nebula is very inhomogeneous in its properties. It is likely that multiple sources of ionization exist within the nebula.

The mid-infrared source in the nebula is very luminous; if the SED is similar to Arp~220, the implied far-infrared luminosity is $L_{\rm FIR}\approx 5.8\times10^{13}~{\rm \Lsun}$. The mid-infrared SED is well modelled as a power-law $\propto \nu^{-2}$, and bears more resemblance to the SEDs of active galaxies and quasars than to star-forming galaxies. We suggest that the MIPS source is a forming, enshrouded, AGN, perhaps similar to the $z\sim 2.5$ extremely luminous dust-enshrouded AGN recently discovered through {\it Spitzer} IRS spectroscopy \citep[]{hou2005}. Spectroscopy with {\it Spitzer} IRS will be necessary to determine the relative contributions of a hot dust continuum component and PAH emission to the total mid-IR flux, and will help us better constrain the relative roles of young stars and AGN in heating the dust.

Since the large \lya\ nebula encompasses a revealed  `Lyman Break galaxy' and an enhrouded active region, we suggest a scenario in which the two objects represent sites of past and present galaxy formation and the \lya\ emission traces the region of current infall. In a manner similar to that observed in nearby HII regions, the galaxy formation ``front'' in this system is moving gradually to the south (perhaps at a rate of $\sim 300\ {\rm kpc \ Gyr^{-1}}$).

We speculate that \lya\ nebulae such as this one may be found through their associated far-infrared emission. Although this may appear contradictory, it is quite plausible that both strong 24\micron\ emisison from dust (powered by enshrouded star-formation and/or accretion activity) and strong \lya\ emission (powered by the cooling of gas in the deep potential well and perhaps by photoionizing radiation escaping from the enshrouded regions) are joint signposts for the spectacular events associated with the formation of the most massive galaxies. 

\acknowledgments

This work is based in part on observations made with the Spitzer Space Telescope, which is operated by the Jet Propulsion Laboratory, California Institute of Technology under NASA contract 1407. We are grateful to the expert assistance of Paola Amico and Joel Aycock at W. M. Keck Observatory and to the staff of Kitt Peak National Observatory where the Bo\"otes field observations of the NDWFS were obtained. The authors thank NOAO for supporting the NOAO Deep Wide-Field Survey. In particular, we thank Lindsey Davis, Alyson Ford, Emma Hogan, Lissa Miller, Erin Ryan, Glenn Tiede and Frank Valdes for their able assistance with the NDWFS data. We also thank Ms. Miranda Nordhaus, Dr. Dianne Harmer, Dr. Richard Green and Dr. George Jacoby for enabling the WIYN observations discussed in this paper. AD thanks Joan Najita, Steve Strom and Joe Shields for useful discussions related to this system. We thank Rob Ivison, Henrik Spoon and Laura Silva for providing template SEDs. We thank an anonymous referee for useful comments. NOAO is operated by the Association of Universities for Research in Astronomy (AURA), Inc. under a cooperative agreement with the National Science Foundation.  The authors also wish to recognize and acknowledge the very significant cultural role and reverence that the summit of Mauna Kea has always had within the indigenous Hawaiian community. 

\bibliography{ms}

\end{document}